\documentclass[nohyperref]{article}

\usepackage{microtype}
\usepackage{graphicx}
\usepackage{subfigure}
\usepackage{booktabs} 

\usepackage{hyperref}

\usepackage[accepted]{icml2023}

\usepackage{amsmath,amsfonts,bm}

\def\eqref#1{equation~\ref{#1}}

\def\1{\bm{1}}

\def\va{{\bm{a}}}
\def\vb{{\bm{b}}}
\def\vc{{\bm{c}}}
\def\vd{{\bm{d}}}

\def\vh{{\bm{h}}}

\def\vk{{\bm{k}}}
\def\vl{{\bm{l}}}

\def\vn{{\bm{n}}}

\def\vp{{\bm{p}}}

\def\vu{{\bm{u}}}
\def\vv{{\bm{v}}}
\def\vw{{\bm{w}}}
\def\vx{{\bm{x}}}

\def\vz{{\bm{z}}}

\def\mA{{\bm{A}}}

\def\mE{{\bm{E}}}

\def\mG{{\bm{G}}}

\def\mL{{\bm{L}}}
\def\mM{{\bm{M}}}

\def\mP{{\bm{P}}}

\DeclareMathAlphabet{\mathsfit}{\encodingdefault}{\sfdefault}{m}{sl}
\SetMathAlphabet{\mathsfit}{bold}{\encodingdefault}{\sfdefault}{bx}{n}

\newcommand{\sigmoid}{\sigma}

\usepackage{amsmath}
\usepackage{amssymb,enumitem}
\usepackage{mathtools}
\usepackage{amsthm}
\usepackage{wrapfig}
\usepackage{graphics}
\usepackage{dsfont}

\usepackage[capitalize,noabbrev]{cleveref}

\theoremstyle{plain}

\theoremstyle{definition}

\theoremstyle{remark}

\usepackage[textsize=tiny]{todonotes}
\icmltitlerunning{Efficient Approximations of Complete Interatomic Potentials for Crystal Property Prediction}

\begin{document}

\twocolumn[
\icmltitle{Efficient Approximations of Complete Interatomic Potentials for Crystal Property Prediction}



\icmlsetsymbol{equal}{*}

\begin{icmlauthorlist}
\icmlauthor{Yuchao Lin}{tamu}
\icmlauthor{Keqiang Yan}{tamu}
\icmlauthor{Youzhi Luo}{tamu}
\icmlauthor{Yi Liu}{fsu}
\icmlauthor{Xiaoning Qian}{tamu,tamuee}
\icmlauthor{Shuiwang Ji}{tamu}
\end{icmlauthorlist}

\icmlaffiliation{tamu}{Department of Computer Science \& Engineering, Texas A\&M University, College Station, TX, USA}
\icmlaffiliation{tamuee}{Department of Electrical \& Computer Engineering, Texas A\&M University, College Station, TX, USA}

\icmlaffiliation{fsu}{Department of Computer Science, Florida State University, Tallahassee, FL, USA}

\icmlcorrespondingauthor{Shuiwang Ji}{sji@tamu.edu}

\icmlkeywords{Machine Learning, ICML}

\vskip 0.3in
]



\printAffiliationsAndNotice{}  

\begin{abstract}
We study property prediction for crystal materials. A crystal structure consists of a minimal unit cell that is repeated infinitely in 3D space. How to accurately represent such repetitive structures in machine learning models remains unresolved. Current methods construct graphs by establishing edges only between nearby nodes, thereby failing to faithfully capture infinite repeating patterns and distant interatomic interactions. In this work, we propose several innovations to overcome these limitations. First, we propose to model physics-principled interatomic potentials directly instead of only using distances as in many existing methods. These potentials include the Coulomb potential, London dispersion potential, and Pauli repulsion potential. Second, we model the complete set of potentials among all atoms, instead of only between nearby atoms as in existing methods. This is enabled by our approximations of infinite potential summations, where we extend the Ewald summation for several potential series approximations with provable error bounds. Finally, we propose to incorporate our computations of complete interatomic potentials into message passing neural networks for representation learning. We perform experiments on the JARVIS and Materials Project benchmarks for evaluation. Results show that the use of interatomic potentials and complete interatomic potentials leads to consistent performance improvements with reasonable computational costs. Our code is publicly available as part of the AIRS library (\url{https://github.com/divelab/AIRS/tree/main/OpenMat/PotNet}).
\end{abstract}

\section{Introduction}

The past decade has witnessed a surge of interests and rapid developments in machine learning for molecular analysis~\citep{duvenaud2015convolutional}. These initial studies mainly focus on the prediction and generation problems of small molecules. To enable computational analyses, molecules need to be featurized in an appropriate mathematical representation form. 
Recently, with the advances of graph neural networks (GNNs)~\citep{gilmer2017neural,battaglia2018relational,gao2021topology,Liu:Protein}, molecules are more commonly represented as graphs in which each node corresponds to an atom, and each edge corresponds to a chemical bond~\citep{stokes2020deep,wang2020advanced}. 
A variety of molecular graph prediction~\citep{stokes2020deep,wang2020advanced} and generation~\citep{shi2019graphaf,jin2018junction,luo2021graphdf} methods have been developed based on 2D molecular graph representations. A key limitation of the 2D graph representations is that the 3D geometries of molecules are not captured, but such information may be critical in many molecular property prediction problems~\citep{hu2021ogb}. To enable the encoding of 3D molecular geometries in GNNs, a series of 3D GNN methods have been developed for prediction~\citep{schutt2017schnet,gasteiger2019directional,liu2022spherical,wang2022comenet} and generation~\citep{liu2022generating,luo2021autoregressive,hoogeboom2022equivariant} problems.
In these 3D graph representations, each node is associated with the corresponding atom's coordinate in 3D space. Geometric information, such as distances between nodes and angles between edges, is used during message passing in GNNs. 
Recently, these methods have been extended to learn representations for proteins~\citep{jing2020learning,wang2023learning}.

Inspired by the success of GNNs on small molecules, 
\cite{xie2018crystal} developed the crystal graph convolutional neural network~(CGCNN) for crystal material property prediction. Different from small molecules and proteins, crystal materials are typically modeled by a minimal unit cell (similar to a small molecule) that is repeated in 3D space with certain directions and step sizes. In theory, the unit cell is repeated infinitely in 3D space, but any real-world material has finite size. However, given that our modeling is at the atomic level, modeling crystal materials as infinite repetitions of unit cells is approximately accurate. Therefore, a key challenge in crystal material modeling is how to accurately capture the infinite-range interatomic interactions resulted from the repetitions of unit cells in 3D space. Current GNN-based crystal property prediction methods construct graphs by creating edges only between atoms within a pre-specified distance threshold~\citep{xie2018crystal,MegNet,GATGNN,CyAtt,ALIGNN}. Thus, they fail to capture interactions between distant atoms explicitly. 

In this work, we propose a new graph deep learning method, \textbf{PotNet}, with several innovations to significantly advance the field of crystal material modeling. First, we propose to model interatomic potentials directly as edge features in PotNet, instead of using distance as in prior methods. These potentials include the Coulomb potential~\citep{west}, London dispersion potential~\citep{london}, and Pauli repulsion potential~\citep{pauli}. Second, a distinguishing feature of PotNet is to model the \textbf{complete} set of potentials among all atoms, instead of only between nearby atoms as in prior methods. This is enabled by our approximations of infinite potential summations with provable error bounds. We further develop efficient algorithms to compute the approximations. Finally, we propose to incorporate our computations of interatomic potentials and complete interatomic potentials into message passing neural networks for representation learning. We performed comprehensive experiments on the JARVIS and Materials Project benchmarks to evaluate our methods. Results show that the use of interatomic potentials and complete interatomic potentials in our methods leads to consistent performance improvements with reasonable computational costs.

\section{Background and Related Work}
\label{background}

\subsection{Crystal Representation and Property Prediction}
\label{crystal}

A crystal structure can be represented as periodic repetitions of unit cells in the three-dimensional~(3D) Euclidean space, where the unit cell contains the smallest repeatable structure of a given crystal. Specifically, let $n$ be the number of atoms in the unit cell, a crystal can be represented as $\mM=(\mA, \mL)$. Here, $\mA=\{\va_i\}_{i=1}^n=\{(\vx_i, \vp_i)\}_{i=1}^n$ describes one of the unit cell structures of $\mM$, where $\vx_i\in\mathbb{R}^b$ and $\vp_i\in\mathbb{R}^3$ denote the $b$-dimensional feature vector and the 3D Cartesian coordinates of the $i$-th atom in the unit cell, respectively. $\mL=[\vl_1,\vl_2,\vl_3] \in \mathbb{R}^{3\times 3}$ is the lattice matrix describing how a unit cell repeats itself in the 3D space. In the complete crystal structure, every atom in a unit cell repeats itself periodically in the 3D space. Specifically, from an arbitrary integer vector $\vk\in\mathbb{Z}^3$ and the unit cell structure $\mA$, we can always obtain another repeated unit cell structure $\mA^\vk=\{\va_i^\vk\}_{i=1}^n=\{(\vx_i^\vk,\vp_i^\vk)\}_{i=1}^n$, where $\vx_i^\vk=\vx_i$, $\vp_i^\vk=\vp_i+\mL\vk$. Hence, the complete crystal structure $\widetilde{\mA}$ of $\mM$ with all unit cells can be described as 
\begin{equation}
\label{eqn:comp_struc}
\widetilde{\mA} = \bigcup_{\vk\in \mathbb{Z}^3} \mA^\vk.
\end{equation}
In this work, we study the problem of crystal property prediction. Our objective is to learn a property prediction model $f: \mM\to y\in\mathbb{R}$ that can predict the property $y$ of the given crystal structure $\mM$. We will focus on predicting the total energy, or other energy-related properties of crystals.

\subsection{Crystal Property Prediction with Interatomic Potentials}
\label{background:potentail}

Most of the classical crystal energy prediction methods are based on interatomic potentials. According to existing studies in physics~\citep{west,eam,born}, the total energy of a crystal structure can be approximated by the summation of interatomic potentials in the crystal. Particularly, the three following categories of interatomic potentials are widely used in crystals, and they can be considered sufficient for accurate energy approximation.

\begin{itemize}[leftmargin=*]

\item \textbf{Coulomb potential} is caused by the electrostatic interaction of two atoms with charges. Coulomb potentials are closely related to ionic bonding and metallic bonding in crystals~\citep{west}. For any two atoms $\va$ and $\vb$, let $z_\va$ and $z_\vb$ denote the number of charges in the atom $\va$ and $\vb$, and let $d(\va,\vb)$ be the Euclidean distance between the atom $\va$ and $\vb$. The Coulomb potential $V\textsubscript{Coulomb}$ is defined as $V\textsubscript{Coulomb}(\va,\vb)=-\frac{z_\va z_\vb e^2_0}{4\pi\epsilon_0 d(\va,\vb)}$. Here $e_0$ is the elementary charge constant, and $\epsilon_0$ is the permittivity constant of free space.

\item \textbf{London dispersion potential} describes the Van der Waals interaction between atoms. It is often considered in energy estimation since its contribution is cumulative over the volume of crystals~\citep{london} and can be sometimes very strong in bulk crystals, such as sulfur and phosphorus. The mathematical form of this potential is described as $V\textsubscript{London}(\va,\vb) = -\epsilon/d^6(\va,\vb)$, where $\epsilon$ is a hyperparameter.

\item \textbf{Pauli repulsion potential} results from the Pauli exclusion principle that generally exists in all crystal structures. The Pauli exclusion principle forces any two atoms to be sufficiently far away from each other so that their electron orbits do not overlap. Such exclusion interactions lead to Pauli repulsion potential with the form of $V\textsubscript{Pauli}(\va,\vb) = e^{-\alpha d(\va,\vb)}$, where $\alpha$ is a hyperparameter~\citep{buckingham,helium}.

\end{itemize}

\subsection{Crystal Property Prediction with Deep Learning}
\label{related}

Physics-based methodologies have long been employed for crystal energy prediction, albeit with a certain degree of specificity. Typically, these methods are highly specific to a particular type of crystal, implying that a single methodology can only deliver precise approximations for one distinct crystal type. Drawing inspiration from the field of physics, Coulomb matrices, as elucidated by~\citep{rupp2012fast, elton2018applying}, assume a pivotal function in the prediction of crystal energy. Nevertheless, their application is constrained, primarily modeling a specific subset of materials, namely, ionic and metallic materials. Moreover, a significant limitation of these matrices is their lack of permutation invariance. Recently, thanks to the advances of deep learning, many studies have been done to develop a general crystal property predictor for a variety of different crystals with powerful deep neural network models. Some studies~\citep{behler2007generalized, crabnet,jha2018elemnet,jha2019irnet,goodall2020predicting} represent crystals as chemical formulas, and adopt sequence models to predict properties from these string representations. However, more recent studies consider crystals as 3D graphs and employ expressive 3D GNNs~\citep{schutt2017schnet,klicpera2020directional,liu2022spherical}, a family of deep neural networks specifically designed for 3D graph-structured data, to crystal representation learning. CGCNN~\citep{xie2018crystal} is the first method that proposes to represent crystals with radius graphs and adopts a graph convolutional network to predict the property from the graph. Based on the pioneering exploration of CGCNN, subsequent studies~\citep{CyAtt,GATGNN,MegNet,ALIGNN,nequip, OMEE2022100491} propose various 3D GNN architectures to achieve more effective crystal representation learning. Particularly, by enhancing the input features with periodic invariance and periodic patterns, Matformer~\citep{yan2022periodic} develops the currently most powerful 3D GNN architecture for crystals and achieves the best crystal property prediction performance.

\section{Method}

Although existing GNN-based methods have achieved impressive performance in crystal property prediction, they struggle in further boosting the performance due to the approximation of interatomic interactions using functional expansions based on distances and failing in capturing complete interatomic interactions. In this section, we present PotNet, a novel crystal representation model that can overcome these limitations of prior methods. Based on the physical modeling of crystal energy, PotNet explicitly uses interatomic potentials and complete interatomic potentials as input features. The complete interatomic potentials are incorporated into the message passing mechanism of graph neural networks and efficiently approximated by an efficient algorithm.

\subsection{Approximating Crystal Energy with Complete Interatomic Potentials}
\label{importance}

According to the density functional theory (DFT) in physics, for any crystal $\mM=(\mA, \mL)$ with the complete structure $\widetilde{\mA}$ defined in Eqn.~(\ref{eqn:comp_struc}), its total energy 
$E(\mM)$ can be approximated by the embedded atom method~\citep{daw1984embedded,eam,eamcovalent,eamionic,riffe2018vibrational} in the form of
\begin{equation}
\label{eqn:eam}
E(\mM) = \frac{1}{2}\sum_{\va\in\mA} \sum_{\vb\ne \va, \vb\in\widetilde{\mA}} V(\va, \vb) + \sum_{\va\in\mA} F(\rho_\va),
\end{equation}
where $V(\va, \vb)$ denotes the interatomic potentials between the atoms $\va$ and $\vb$, capturing the magnitude of interactions; $\rho_\va$ is the local electron density of the atom $\va$,  determined by the coordinate and number of charges of the atom $\va$ according to the Hohenberg-Kohn theorem; $F(\cdot)$ is a parametrized function to embed the electron density $\rho_\va$. Actually, existing studies~\citep{density} show that $\rho_\va$ can be considered as a function of $\sum_{\vb\ne \va, \vb\in\widetilde{\mA}} V(\va, \vb)$ mathematically\footnote{Under zeroth-order approximation, the electron density $\rho_\va$ is represented as the aggregate of functions, analogous in type to those used for potential energy calculations. Due to computational efficiency, the approximation form of $e^{-\Vert \mL \vk + \vv \Vert^2}$ is intentionally excluded from this study. This series type can be computed using the Riemann Theta Function as described in Appendix~\ref{gls}.}. Hence, Eqn.~(\ref{eqn:eam}) can be rewritten in the following form:
\begin{multline}
\label{eqn:new_eam}
    E(\mM)=\sum_{\va\in\mA}\Bigg[\frac{1}{2}\sum_{\vb\ne \va, \vb\in\widetilde{\mA}}V(\va,\vb)\\
    + G\left(\sum_{\vb\ne \va, \vb\in\widetilde{\mA}}V(\va,\vb)\right)\Bigg],
\end{multline}
where $G(\cdot)$ is a parametrized function. Eqn.~(\ref{eqn:new_eam}) can be considered as a way to compute the energy from the complete interatomic potential summation $\sum_{\vb\ne \va, \vb\in\widetilde{\mA}}V(\va,\vb)$ of every atom $\va$ in the unit cell $\mA$. However, in practice, the function $G$ is computationally expensive if not infeasible. Hence, more and more studies have turned to the powerful learning capability of modern deep neural network models to approximate it effectively. 

\subsection{Limitations of Existing Deep Learning Methods}
\label{sec:limitation}

Currently, most of the existing graph deep learning methods for crystals~\citep{xie2018crystal, MegNet, GATGNN, ALIGNN} use radius graph representations and distance-based features as inputs to predict the crystal energy in Eqn.~(\ref{eqn:new_eam}). Specifically, for a crystal $\mM=(\mA,\mL)$, the radius graph is constructed by adding edges between any atom $\va$ in the unit cell $\mA$ and any other atom $\vb$ in the complete crystal structure $\widetilde{\mA}$ whose distances are smaller than a pre-specified distance threshold $r$. In addition, some functional expansions of distances, e.g., radial basis functions~(RBF), are used to model interatomic interactions and form the input edge features to 3D GNN models. Hence, let $\va=(\vx_\va,\vp_\va), \vb=(\vx_\vb,\vp_\vb)$, the crystal energy prediction $\hat{E}(\mM)$ of these methods can be generally described as
\begin{equation}
    \label{eqn:old_method}
    \hat{E}(\mM)=\sum_{\va\in\mA}\sum_{\vb\in\mathcal{N}_r(\va)} H\left(\phi\left(||\vp_\va-\vp_\vb||_2\right)\right),
\end{equation}
where $\mathcal{N}_r(\va)=\{\vb: \vb\ne \va, \vb\in\widetilde{\mA},  ||\vp_\va-\vp_\vb||_2<r\}$, $\phi(\cdot)$ denotes the functional expansions, and $H(\cdot)$ is a non-linear function based on 3D GNN models.

However, we argue that predicting or approximating the energy with Eqn.~(\ref{eqn:old_method}) is a suboptimal solution. Actually, compared with Eqn.~(\ref{eqn:new_eam}), which is physics-principled, there exist non-negligible approximation errors in Eqn.~(\ref{eqn:old_method}). First, Eqn.~(\ref{eqn:old_method}) captures the interatomic interactions based on interatomic distances, while the energy can be more accurately approximated by a function of interatomic potentials as in Eqn.~(\ref{eqn:new_eam}). Though according to Sec.~\ref{background:potentail}, interatomic potentials themselves are also functions of distances, we argue that directly using functional expansions of distances is not the best solution to crystal energy prediction. The commonly used functional expansions in existing methods, such as RBF $\phi(\cdot)$, have different mathematical forms from potentials defined in Sec.~\ref{background:potentail}. Intuitively, this poses more challenges to 3D GNN models since they need to learn a mapping from $\phi(\cdot)$ to the energy $\mE$, while the energy $\mE$ is not a direct function of $\phi(\cdot)$. Hence, we argue that directly employing the physics-principled potential functions instead of $\phi(\cdot)$ as input features is more suitable for crystal energy prediction. 

Second, different from Eqn.~(\ref{eqn:new_eam}), Eqn.~(\ref{eqn:old_method}) does not capture the complete set of interatomic interactions because the summation set $\mathcal{N}_r(\va)$ of atoms $\vb$ is constrained to be the atoms whose distances to the atom $\va$ are smaller than $r$. This can lead to an approximation error due to ignoring the accumulation of interatomic potentials. By the first principles in physics, interatomic potentials decay algebraically when pairwise interatomic distances become larger. Hence, for a finite structure like molecules, the potentials from atoms that are far away from a given atom are limited and can be ignored. However, this cannot be ignored for crystals since they accumulate infinitely. As a result, the accumulation of interatomic potentials can have a significant effect on a given atom in the infinite crystal structure. Let $d$ be the distance between atoms $\va$ and $\vb$ and $p$ be a positive real number. Considering interatomic potentials $V\propto 1/d^p$, and assuming a 3D crystal structure containing an atom repeating itself with a Euclidean distance of 1, then the energy contribution by considering all its repetitions to it is simply the summation of all these interatomic potentials. To be concrete, the total potential summation $\widetilde{V}$ satisfies $\widetilde{V}\propto \sum_{\vk\in\mathbb{Z}^3} 1/\Vert\vk\Vert^p$. Considering potentials of the pairwise atoms within the distance threshold $r$, i.e., a sphere $S_r$, we have the smallest possible prediction error $\Delta V$ satisfying $\Delta V \propto \sum_{\vk\in\mathbb{Z}^3/S_r } 1/\Vert\vk\Vert^p$. Different from the geometry series, $\Delta V$ decays at an approximately algebraical rate rather than exponentially. This suggests that a large radius $r$ is needed to accurately approximate $\widetilde{V}$. Taking London dispersion potential $p=6$ as an example and it can be calculated that to approximate $\sum_{\vk\in\mathbb{Z}^3} 1/\Vert\vk\Vert^6\approx 8.40$ with $0.01$ absolute error, we need at least $r = 56$, while in common radius crystal graph construction~\citep{xie2018crystal,MegNet,ALIGNN,GATGNN,schutt2017schnet}, the radius covers only a unit cell and its neighbors at average (see Appendix~\ref{sec:dataset}), analogy to $r=1$. In addition, a larger radius will consume much more time for crystal graph construction since it induces a cubic time complexity in the 3D space.
We can observe from this example that the failure to capture complete interatomic potentials due to the use of radius is a key factor that prevents accurate energy prediction in existing GNN-based methods. In addition, we experimentally show that large cutoffs will produce better results for classic crystal-graph-based GNNs but consume more processing time in Appendix~\ref{sec:additional}. To remedy this problem, our efficient algorithm is presented in Sec.~\ref{algorithm}.


\begin{figure*}[t]
    \centering
    \includegraphics[width=\linewidth]{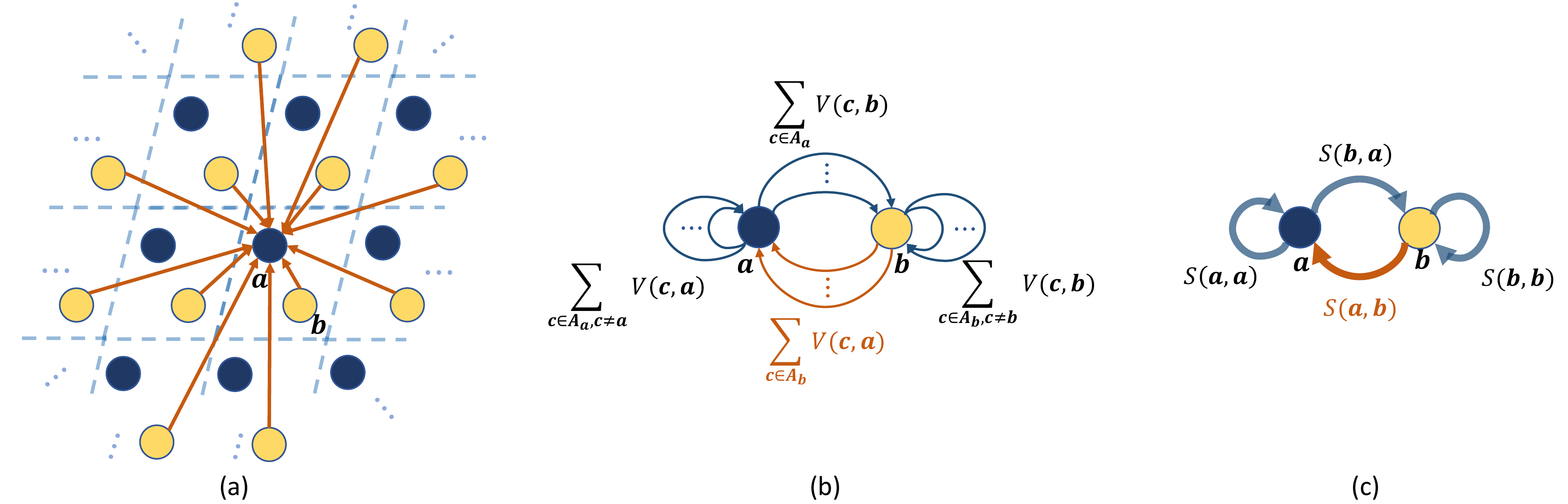}
    \vskip -0.1in
    \caption{
    Schematic illustrations of how complete interatomic interactions are captured in PotNet. Note that PotNet models 3D crystals while we have 2D illustration for simplicity. (a) An example crystal in which each unit cell contains two atoms $a$ and $b$. In PotNet, the potentials between all pairs of atoms are captured. For simplicity, we only show the potentials from all $b$ atoms to a $a$ atom. (b) The complete set of potentials in (a) can be grouped into four categories, including $a\rightarrow b$, $b\rightarrow a$, $a\rightarrow a$, and $b\rightarrow b$. (c) We propose to compute an approximate summation for each category of potentials. 
    }
    \label{fig:infinite}
\vspace{-0.4cm}\end{figure*}

\subsection{Message Passing with Complete Interatomic Potentials}
\label{icg}

It follows from the analysis in Sec.~\ref{sec:limitation} that major limitations of existing deep learning methods for crystal representation learning lie in: (a) not making predictions from physics-principled interatomic potentials, and (b) not considering complete interatomic interactions. To overcome these limitations, we propose to explicitly use complete interatomic potential summations in GNN models. Since our proposed method is tightly related to potentials, we name it PotNet.

By reformulating Eqn.~(\ref{eqn:new_eam}), our PotNet incorporates the crystal energy computation with complete interatomic potentials into the message passing scheme of GNN models. For any material structure $\mM=(\mA,\mL)$, we can rewrite the definition of its complete structure $\widetilde{\mA}$ in Eqn.~(\ref{eqn:comp_struc}) as    
\begin{multline}
    \label{eqn:new_comp_struc}
    \widetilde{\mA} = \bigcup_{\vk\in \mathbb{Z}^3} \mA^\vk=\bigcup_{\vk\in\mathbb{Z}^3}\bigcup_{\vb\in\mA}\{\vb^\vk\}=\bigcup_{\vb\in\mA}\bigcup_{\vk\in\mathbb{Z}^3}\{\vb^\vk\}\\
    =\bigcup_{\vb\in\mA}\mA_\vb,
\end{multline}
where $\mA_\vb=\bigcup_{\vk\in\mathbb{Z}^3}\{\vb^\vk\}$ denotes the set of atoms containing the atom $\vb$ from the unit cell $\mA$ and all its periodic repetitions in the complete crystal structure. With Eqn.~(\ref{eqn:new_comp_struc}), we can reformulate Eqn.~(\ref{eqn:new_eam}) as
\begin{equation}
\label{eqn:graph_eam}
  \begin{split}
    E(\mM)&=\begin{multlined}[t] \sum_{\va\in\mA}\Bigg[\frac{1}{2}\sum_{\vb\in\mA}\sum_{\vc\ne\va,\vc\in\mA_\vb}V(\va,\vc)\\
    +G\left(\sum_{\vb\in\mA}\sum_{\vc\ne\va,\vc\in\mA_\vb}V(\va,\vc)\right)\Bigg]\end{multlined}\\
    &=\sum_{\va\in\mA}\left[\frac{1}{2}\sum_{\vb\in\mA}S(\va,\vb)+G\left(\sum_{\vb\in\mA}S(\va,\vb)\right)\right],
  \end{split}
\end{equation}
where the infinite potential summation $S(\va,\vb)=\sum_{\vc\ne\va,\vc\in\mA_\vb}V(\va,\vc)$ denotes the sum of the interatomic potentials from the atom $\vb$ together with its all periodic repetitions to the atom $\va$. Eqn.~(\ref{eqn:graph_eam}) can be integrated into the message passing scheme of GNN models. Specifically, we can create a graph $\mG$ for $\mM=(\mA,\mL)$, where each atom in the unit cell $\mA$ corresponds to a node in the graph. For any two nodes $\vu,\vv$ in the graph, there is an edge connecting them, and every node $\vu$ in the graph is also connected to itself by a self-loop edge. If we consider the infinite potential summation $S(\va,\vb)$ as the feature of the edge from node $\vb$ to $\va$, we can use the message passing based non-linear neural network model in GNN to fit the function $\frac{1}{2}\sum_{\vb\in\mA}S(\va,\vb)+G\left(\sum_{\vb\in\mA}S(\va,\vb)\right)$.

\begin{table*}[t]
\caption{Comparison between our method and other baselines in terms of test MAE on the Materials Project dataset. To make the comparison clear and fair, we follow~\citet{yan2022periodic} and use the same dataset settings. The best results are shown in \textbf{bold} and the second best results are shown with \underline{underlines}.}
\label{tb:mp}
\begin{center}
\begin{tabular}{l|cccc}
\toprule
& Formation Energy & Band Gap & Bulk Moduli & Shear Moduli \\
\cmidrule(r){2-5}
Method & eV/atom  &  eV &  log(GPa) & log(GPa)  \\
\midrule
CGCNN & 0.031 & 0.292  & 0.047 &0.077 \\
SchNet & 0.033 & 0.345 & 0.066 & 0.099 \\
MEGNET & 0.030 & 0.307 & 0.051 & 0.099 \\
GATGNN & 0.033 & 0.280 & 0.045 & 0.075 \\
ALIGNN & 0.0221 & 0.218 & 0.051 & 0.078 \\
Matformer & \underline{0.0210} & \underline{0.211} & \underline{0.043} & \underline{0.073} \\
PotNet  & \textbf{0.0188} & \textbf{0.204} & \textbf{0.040} & \textbf{0.065} \\
\bottomrule
\end{tabular}
\end{center}
\vskip -0.25in
\end{table*}

Based on this design of directly using interatomic potentials as edge features, our PotNet employs a GNN model with multiple message passing layers on the graph $\mG$ to predict the crystal energy of $\mM$. The computational process of the $\ell$-th message passing layer for the node $\va$ can be described as
\begin{equation}
\label{eqn:mpnn}
    \vh_\va^{(\ell)} = g_\varphi\left(\vh_\va^{(\ell-1)}, \sum_{\vb\in \mA} f_\theta\left(\vh_\va^{(\ell-1)}, \vh_\vb^{(\ell-1)}, S(\va,\vb)\right)\right),
\end{equation}
where $\vh_\va^{(\ell)}$ denotes the embedding vector of node $\va$ generated from the $\ell$-th message passing layer, $\vh_\va^{(0)}$ is initialized to the atom feature vector of the atom $\va$, and $g_\varphi(\cdot), f_\theta(\cdot)$ are both neural network models with trainable parameters $\varphi$ and $\theta$, respectively. Here, the model $f_\theta$ plays the role of capturing information from both atomic features and complete interatomic potentials. Detailed information about model architectures of $f_\theta$ and $g_\varphi$ is provided in Appendix~\ref{model}. Note that our PotNet is actually a 3D GNN model even though 3D geometric information is not explicitly involved in Eqn.~(\ref{eqn:mpnn}). This is because the edge feature $S(\va,\vb)$ is related to potential functions, and by Sec.~\ref{background:potentail} we know that they are computed from interatomic distances. In other words, PotNet can be considered to encode 3D geometric information with potential functions, though our direct motivation of using potential functions comes from the physical modeling of crystal energy.

Intuitively, the message passing process in Eqn.~(\ref{eqn:mpnn}) over the graph $\mG$ can be considered as a general case of employing a radius graph where the distance threshold $r$ goes to infinity, i.e., $r\to+\infty$. In this case, as shown in Fig.~\ref{fig:infinite}(a), for any atom in the crystal, all the other atoms in the complete crystal structure have been included to interact with it. If we follow the radius graph construction process in the previous methods~\citep{xie2018crystal, MegNet, GATGNN, ALIGNN}, we obtain a graph $\widetilde{\mG}$ in which there exist an infinite number of edges between every pair of nodes. However, PotNet simplifies this complicated graph $\widetilde{\mG}$ to the graph $\mG$ in which only one edge exists between every node pair. Specifically, PotNet directly models interatomic interactions as potentials and for any two nodes in $\widetilde{\mG}$, PotNet aggregates all edges between them to a single edge by the use of infinite potential summation $S(\va,\vb)$ (see Fig.~\ref{fig:infinite}(b)). In other words, PotNet provides an effective solution that enables GNN models to capture complete interatomic interactions through the use of infinite potential summations.

\subsection{Efficient Computation of Infinite Potential Summation}
\label{algorithm}

Although the infinite potential summations have been effectively incorporated into the message passing based GNN models, the computation of these infinite potential summations is not trivial. Basically, there are two challenges to achieve accurate and efficient computation of the infinite potential summations. For accuracy, the computation algorithm requires provable error bounds. For efficiency, fast algorithm is needed to achieve scalable GNN training and fast crystal property prediction. To tackle these two challenges, we derive a fast approximation algorithm for infinite potential summations based on the Ewald summation method~\citep{ewald, kosmala2023ewald}. To be concrete, we unify the summations of three infinite potentials between the position of atom $\va$ and all repeated positions of atom $\vb$ into an integral form, so that the Ewald summation method can be efficiently implemented in PotNet~(Fig.~\ref{fig:infinite}(c)). The key idea of the Ewald summation is that a slowly converging summation in the real space is guaranteed to be converted into a quickly converging summation in the Fourier space~\citep{woodward2014probability, kosmala2023ewald}. Based on this, the Ewald summation method divides a summation into two parts. One part has a quicker converging rate in the real space than the original summation. The other slower-to-converge part is then transformed into the Fourier space and becomes quickly convergent. In our method, the Ewald summation method is used with the infinite summations by dividing the integral into two parts, including one part that converges quickly in the Fourier space and another part that converges quickly in the real space, to obtain a fast approximation with provable error bounds.

\begin{table*}[t]
\caption{Comparison between our method and other baselines in terms of test MAE on JARVIS dataset. The best results are shown in \textbf{bold} and the second best results are shown with \underline{underlines}.}
\label{tb:jarvis}
\begin{center}
\begin{tabular}{l|ccccc}
\toprule
& Formation Energy & Bandgap(OPT) & Total energy & Bandgap(MBJ) & Ehull\\ 

\cmidrule(r){2-6}
Method & eV/atom  &  eV & eV/atom & eV & eV   \\
\midrule
CFID & 0.14 &    0.30 & 0.24  &  0.53 & 0.22 \\
CGCNN  & 0.063 &   0.20 & 0.078 &  0.41 & 0.17  \\
SchNet & 0.045 &   0.19 & 0.047 & 0.43 & 0.14  \\
MEGNET  & 0.047 &   0.145 & 0.058 & 0.34 & 0.084  \\
GATGNN  & 0.047 &   0.17 & 0.056 & 0.51 & 0.12  \\
ALIGNN  & 0.0331 & 0.142 & 0.037 & 0.31 & 0.076\\
Matformer  & \underline{0.0325} & \underline{0.137} & \underline{0.035} & \underline{0.30} & \underline{0.064}\\
PotNet   & \textbf{0.0294} & \textbf{0.127} & \textbf{0.032} & \textbf{0.27} & \textbf{0.055}\\
\bottomrule
\end{tabular}
\end{center}
\vskip -0.25in
\end{table*}

Following notations in Sec.~\ref{background} and Sec.~\ref{icg}, we denote the positions of atoms in the set $\mA_\vb$ as $\mP_\vb=\{\vp_\vb^\vk\,|\, \vp_\vb^\vk=\vp_\vb+\mL\vk,\vk\in\mathbb{Z}^3\}$. The Euclidean distances between the atom $\va$ and all atoms in $\mA_\vb$ can be represented as $\{d\,|\,d = \Vert \vp_\vb + \mL\vk - \vp_\va \Vert, \vk\in \mathbb{Z}^3\}$. We then investigate the three types of potential mentioned in Sec.~\ref{background:potentail}. Since charges and constants in the Coulomb potential function can be extracted outside the summation and modeled as part of atom features, we simplify the Coulomb potential function as $V\textsubscript{Coulomb}(\va,\vb) = -\epsilon^{\prime}/d(\va,\vb)$, where $\epsilon^{\prime}$ is a hyperparameter scaling the Coulomb potential. As a result, we can represent Coulomb potentials from all atoms in $\mA_\vb$ to the atom $\va$ as $\{-\frac{\epsilon^{\prime}}{d} \,\,|\,\,d = \Vert \vp_\vb + \mL\vk - \vp_\va \Vert, d\ne 0 ,\vk\in \mathbb{Z}^3\}$. Similarly, London dispersion potentials from all atoms in $\mA_\vb$ to the atom $\va$ can be represented as $\{-\frac{\epsilon}{d^6} \,\,|\,\,d = \Vert \vp_\vb + \mL\vk - \vp_\va \Vert, d\ne 0, \vk\in \mathbb{Z}^3\}$. It is worth noting that Coulomb and London dispersion potentials can be represented in a unified view as $\{\frac{\text{constant}}{d^p} \,\,|\,\,d = \Vert \vv_{\va\vb} + \mL\vk \Vert, d\ne 0, \vv_{\va\vb} = \vp_\vb - \vp_\va, \vk\in \mathbb{Z}^3\}$ with a positive real number $p$. In addition, we represent Pauli potentials from all atoms in $\mA_\vb$ to the atom $\va$ as $\{e^{-\alpha d} \,\,|\,\,d = \Vert \vv_{\va\vb} + \mL\vk \Vert, \vv_{\va\vb} = \vp_\vb - \vp_\va, \vk\in \mathbb{Z}^3\}$ with a hyperparameter $\alpha$. We provide detailed proofs in Appendix~\ref{proof_34_1} that the summations of these three potentials can be unified in an integral form as
\vspace{-0.1cm}
\begin{multline}
\label{equ:sum}
    S(\va, \vb) = D \int_0^\infty t^{C-1}(-\delta(\vv, B) +\\
    \sum_{\vk \in \mathbb{Z}^3} e^{-A\pi \Vert\mL\vk+\vv_{\va\vb}\Vert^{2}t - \frac{B}{t}})dt,
\end{multline}
where $A,B,C,D$ are constants derived from the corresponding specific potential forms and $\delta$ is the generalized delta function such that $\delta(\vv, B) = 1$ if and only if $\vv = \mathbf{0}$ and $B = 0$, otherwise $\delta(\vv, B) = 0$. We then apply the Ewald summation method~\citep{ewald} to Eqn.~(\ref{equ:sum}) and split it into two parts as
\begin{equation}
\begin{aligned}
    S(\va, \vb) = S\textsubscript{Fourier}(\va, \vb) + S\textsubscript{direct}(\va, \vb), 
\end{aligned}
\end{equation}
where $S\textsubscript{direct}$ denotes the short-range part that converges quickly in real space, $S\textsubscript{Fourier}$ denotes the long-range\footnote{Note that the distinction between the short-range and long-range terms is not determined by a radius cutoff in Euclidean space. Instead, it is delineated by the rate of series convergence, or in more mathematical terms, a constant point in the integral summation. This can be represented as $\int_0^1 \sum \cdot\ dt + \int_1^\infty\sum \cdot \ dt$, where $\int_0^1 \sum \cdot\ dt$ denotes the series with a slower convergence rate in the direct space but a faster rate in the Fourier space. Conversely, $\int_1^\infty \sum \cdot\ dt$ denotes the opposite scenario.} part that converges quickly in Fourier space, and the total summation converges as shown by \citet{ewald}. We demonstrate in Appendix~\ref{integralsummation} that $S\textsubscript{direct}$ and $S\textsubscript{Fourier}$ can be represented as sums of incomplete Bessel functions $K_\nu (x,y)$. Rigorous mathematical proofs are supplied in~\ref{besselconverge}, establishing the convergence of these summations of incomplete Bessel functions and their ability to be approximated with an error that remains within the bounds set by the Gaussian Lattice Sum. The practical application of the proposed summation algorithm is outlined in detail in Appendix~\ref{example}. 

Notably, the summations of incomplete Bessel functions pertaining to London dispersion potentials and Pauli potentials can be approximated directly. In contrast, the summations related to Coulomb potentials are computed by leveraging established mathematical work~\citep{terras1973bessel,kirsten1994generalized}, and employing analytic continuation to extend the domain of $p$, as elucidated in Appendix~\ref{analytic} and~\ref{sec:zeta}. Considering the crystal system's inherent tendency towards neutrality and equilibrium, we illustrate the application of Coulomb potential summation in Appendix~\ref{born-lande}. It is of significance to note that PotNet stands as the pioneering methodology to apply the incomplete Bessel function for computing the Pauli potential summation, a feat unattainable by previous methods~\citep{epstein, lee2009ewald, fft}. Additionally, our method enables computation of other types of interatomic potential summations, including the Lennard-Jones potential, Morse potential, and screened Coulomb potential, as detailed in Appendix~\ref{addition}.

\section{Experimental Studies}
\label{experiments} 

\subsection{Experimental Setup} \label{sec:setup}

We conduct experiments on two material benchmark datasets, including The Materials Project and JARVIS. Baseline methods include CFID~\citep{CFID}, SchNet~\citep{schutt2017schnet}, CGCNN~\citep{xie2018crystal}, MEGNET~\citep{MegNet}, GATGNN~\citep{GATGNN}, ALIGNN~\citep{ALIGNN}, and Matformer~\citep{yan2022periodic}. All PotNet models are trained using the Adam~\citep{kingma2014adam} optimizer with weight decay~\citep{loshchilov2017decoupled} and one cycle learning rate scheduler~\citep{smith2018super} with a learning rate of 0.001, training epoch of 500, and batch size of 64. We use Pytorch and Cython to implement our models. For all tasks on two benchmark datasets, we use one NVIDIA RTX A6000 48G GPU as well as Intel Xeon Gold 6258R CPU for computing. Other detailed configurations of PotNet are provided in Appendix~\ref{model}.

In the implementation, PotNet employs both local and infinite crystal graphs, allowing it to encapsulate global, infinite interactions without compromising the fidelity of local interactions. More specifically, for the local crystal graph, we adopt the radius crystal graph as introduced by CGCNN~\citep{xie2018crystal}, albeit with modifications; we substitute Euclidean distances with interatomic potentials to serve as edge features. Given that the impacts of both London dispersion and Pauli potentials are circumscribed within the radius crystal graph, and can be largely disregarded when focusing solely on radius regions, we restrict ourselves to using Coulomb potentials within the radius crystal graph. In contrast, the infinite crystal graph is formulated as detailed in Section~\ref{icg}, where we incorporate all forms of interatomic potentials - Coulomb, London dispersion, and Pauli potentials. This approach enables us to capture both local and global interactions simultaneously, increasing the model's capacity for crystal representation. In order to enhance our model by leveraging infinite potential features, we have also integrated two supplementary techniques, namely the incorporation of periodic table information and the implementation of transformer operations~\citep{graphormer, yan2022periodic}. It should be noted that these methods are not applied in the main body of this paper. They have been exclusively applied and discussed in the supplementary section, as elucidated in Appendix~\ref{sec:improve}.

\begin{table}[t]
\vskip -0.1in
\caption{Training time per epoch, total training time, total testing time, and model complexity compared with ALIGNN and Matformer on JARVIS formation energy prediction.}
\begin{center}
\vskip -0.1in
\label{tb:efficiency}
\resizebox{\columnwidth}{!}{
\begin{tabular}{l|cccc}
\toprule
Method   & Time/Epoch & Total Training Time & Total Testing Time & Model Para. \\ 
\midrule
ALIGNN & 327 s &  27.3 h & 156 s & 15.4MB  \\
Matformer & 64 s &  8.9 h & 59 s & 11.0 MB  \\
Ours  & \textbf{42 s} &  \textbf{5.8 h} & \textbf{31 s} & \textbf{6.7 MB} \\
\bottomrule
\end{tabular}
}
\end{center}
\vskip -0.3in
\end{table}

\subsection{Experimental Results}
\label{exp_results}

\textbf{The Materials Project}. We first evaluate PotNet on The Materials Project-2018.6.1, which is a widely used large-scale material benchmark with 69239 crystals. Following previous works~\citep{xie2018crystal,MegNet,ALIGNN,GATGNN,yan2022periodic}, four crystal properties including formation energy, band gap, bulk moduli, and shear moduli, are used for evaluating our model. We notice that previous works~\citep{xie2018crystal,MegNet,ALIGNN,GATGNN,schutt2017schnet} compare with each other using different versions of splitting training, evaluation, and testing datasets with different random seeds. For instance, \emph{the original CGCNN paper only uses 28046 training samples for formation energy prediction, resulting in the original result of 0.039.} To make the comparisons fair, we follow the settings of the previous state-of-the-art~(SOTA) Matformer~\citep{yan2022periodic} for all tasks since they retrain all baselines using the same dataset settings and the same data splits from ALIGNN~\citep{ALIGNN}. Note that most of the retrained results recorded in \citet{yan2022periodic} are better than their counterparts in their original papers. We present our results in Table~\ref{tb:mp}, where PotNet consistently outperforms other SOTA methods on all four tasks. 
Furthermore, the impressive results of PotNet for Bulk Moduli and Shear Moduli tasks with only 4664 training samples demonstrate the robustness and adaptive ability of PotNet to the tasks with small training data.  

\textbf{JARVIS Dataset}. We then evaluate PotNet on JARVIS-DFT-2021.8.18 3D dataset, a newly released benchmark dataset proposed by \citet{choudhary2020joint} with 55722 crystals. We evaluate PotNet on five crystal property prediction tasks, including formation energy, bandgap~(OPT), bandgap~(MBJ), total energy, and Ehull. We follow Matformer~\citep{yan2022periodic} and use the same training, validation, and test splits for all these tasks, and also use their retrained baseline results. As shown in Table~\ref{tb:jarvis}, PotNet achieves the best performances on all five tasks consistently. The superior performances of PotNet show the effectiveness of interaction modeling using potentials and explicit modeling of infinite interactions for crystal structures. 


\textbf{Efficiency of PotNet}. Beyond the superior modeling capacity for crystals, our PotNet is faster and more efficient than ALIGNN and Matformer. To demonstrate the efficiency of PotNet, we compare PotNet with ALIGNN and Matformer in terms of training time per epoch, total training time, inference time and model parameters for the task of JARVIS formation energy prediction. From Table~\ref{tb:efficiency},  PotNet is four times faster in terms of total training time and inference time compared with ALIGNN, and also faster than Matformer by 34\% and 47\% in terms of training time per epoch and inference time, respectively.

We also analyze the time cost of the infinite potential summation algorithm in Table~\ref{table:algorithm_cost}. Since there lack baselines for comparison of our infinite potential summation, we compare the prediction time (involving both data preprocessing and model inference time) with the most recent methods ALIGNN and Matformer. To perform preprocessing, unlike previous methods, we need to compute the infinite potential summations besides constructing graphs. However, as shown in Table~\ref{table:algorithm_cost}, the mean prediction time of PotNet for a single crystal is at the level of milliseconds, which is similar to ALIGNN and Matformer. Particularly, PotNet has a faster prediction speed than ALIGNN as the latter involves the computing of angles. 
PotNet is slightly slower than Matformer even though our PotNet needs to compute infinite summations, yet PotNet is much more effective.
To provide more details, we further present numerical examples and their corresponding time cost of our infinite potential summation algorithm in Table~\ref{table:algorithm} in Appendix~\ref{example}.


\begin{table}[t]
\caption{Prediction time cost of infinite potential summation algorithm on the JARVIS test dataset with 5572 crystals. The prediction time considers both preprocessing and model inference time.}
\begin{center}
\vspace{-0.4cm}

\label{table:algorithm_cost}
\resizebox{\columnwidth}{!}{
\begin{tabular}{l|cc}
\toprule
Method & Total Prediction Time & Prediction Time/Per Crystal\\
\midrule
 ALIGNN & 167s & 30 ms\\
 Matformer & 67s  & 12ms\\
 PotNet & 91s  & 16ms\\
\bottomrule
\end{tabular}}
\end{center}
\vspace{-0.8cm}
\end{table}

\subsection{Ablation Studies}

In this section, we demonstrate the importance of two core components of PotNet, including interaction modeling using interatomic potentials and infinite potential summations for crystal prediction. We conduct experiments on the JARVIS-DFT 3D formation energy task, and use test MAE for evaluation. In order to highlight the infinite potential summations, a comprehensive ablation study pertaining to various metrics, performed on the JARVIS-DFT 3D dataset, is detailed in Appendix~\ref{sec:full_ablation}.

\textbf{Interaction Modeling using Potentials}. We demonstrate the importance of interaction modeling using potentials by directly replacing potentials with Euclidean distances used by previous works in our PotNet with exactly the same model architecture. Specifically, we denote PotNet with only local crystal graph as the base model. We use `Base + Euclidean' to represent the base model with Euclidean distances as edge features and `Base + Potential' to represent the base model using Coulomb potentials as edge features. As shown in Table~\ref{ablation}, by replacing Euclidean distances with Coulomb potentials, PotNet without considering infinite potential summation already obtains a significant performance gain from 0.0363 to 0.0301, revealing the importance of interaction modeling using potentials in PotNet.

\textbf{Infinite Potential Summations}. The importance of infinite summation of potentials is demonstrated by comparing the previous base models with `Base + Potential + Infinite', denoting the full PotNet model with infinite summation in infinite crystal graph. It can be seen from Table~\ref{ablation} that by using infinite crystal graphs, the global information of crystal structures is captured, resulting in a performance gain from 0.0301 to 0.0294 for formation energy prediction.

\section{Limitation}

Our research indicates an encouraging performance enhancement using interatomic potentials and complete interatomic potentials. However, we must acknowledge a limitation: the interatomic potential's inherent incapacity to account for interactions extending beyond two atomic participants. Even though the total potential of a material system can be estimated by the embedded atom method~\citep{eam}, which sidesteps the necessity for many-body interactions, explicit consideration of angular and many-body interactions may lead to more precise simulations. This is due to the potential influence of additional atoms on interatomic interactions, which could subsequently modify the total potential outcome. Future studies might gain from integrating many-body interactions into the model, such as modeling the three-body potential and extending this methodology to an infinite summation approach.

\begin{table}[t]
\caption{Ablation studies for the effects of adding Coulomb potentials and infinite summation.}
\begin{center}
\vspace{-0.3cm}
\label{ablation}
\resizebox{\columnwidth}{!}{
\begin{tabular}{l|c}
\toprule
  & JARVIS Formation Energy \\ 
\cmidrule(r){2-2}
Method  & eV/atom \\ 

\midrule
Base + Euclidean        & 0.0363\\
Base + Potential             & 0.0301 \\
Base + Potential + Infinite\quad   \quad \quad \quad           & \textbf{0.0294} \\
\bottomrule
\end{tabular}}
\end{center}
\vspace{-0.9cm}
\end{table}

\section{Conclusion}

We study the problem of how to capture infinite-range interatomic potentials in crystal property prediction directly. As a radical departure from prior methods that only consider distances among nearby atoms, we develop a new GNN, PotNet, with a message passing scheme that considers interatomic potentials as well as efficient approximations to capture the complete set of potentials among all atoms. Experiments show that the use of complete potentials leads to consistent performance improvements. Altogether, our work provides a theoretically principled and practically effective framework for crystal modeling. In the future, we expect our approximations may be further improved to obtain a lower error bound. We also expect our algorithms for computing the summation could be further improved.

\section*{Acknowledgements}

This work was supported in part by National Science Foundation grants IIS-1908220, CCF-1553281, IIS-1812641, DMR-2119103, and IIS-2212419, and National Institutes of Health grant U01AG070112.

\bibliography{potnet,dive,potnet_ref}
\bibliographystyle{icml2023}

\newpage
\appendix
\onecolumn

\section{Gaussian Lattice Sum}
\label{gls}
Let $\mL \in \mathbb{R}^{d\times d}$ be the full rank lattice matrix and $\vv \in {\mathbb{R}^d}$. By extending the definition of \textit{Gaussian Lattice Sum}~\citep{gaussian}, we consider a summation of $d$-dimensional Gaussian functions on a given shifted lattice,
\begin{align}
E(\mL, \vv, \alpha) = \sum_{\vk \in \mathbb{Z}^d} e^{-\pi \alpha\Vert \mL \vk + \vv\Vert^2},\quad \alpha > 0.
\end{align}
One of the characteristics of the Gaussian Lattice Sum is the term $e^{-\pi \alpha\Vert \mL \vk + \vv\Vert^2}$ rapidly decays as $\vk$ becomes large, leading to a fast converging rate of $E(\mL, \vv, \alpha)$. By simple derivation, Gaussian Lattice Sum can be written w.r.t \textit{Riemann Theta Function},
\begin{equation}
E(\mL, \vv, \alpha) = \sum_{\vk \in \mathbb{Z}^d} e^{-\pi \alpha\Vert \mL \vk + \vv\Vert^2} = \theta \left(i\alpha\mL^T \vv | i\alpha \mL^T \mL\right) e^{-\pi \alpha \vv^T \vv},
\end{equation}
where $\theta( \vz | \mathbf{\Omega}) = \sum_{\vk \in \mathbb{Z}^d} e^{2\pi i (\frac{1}{2} \vk \cdot \mathbf{\Omega} \cdot \vk + \vk \vz)}$ denotes \textit{Riemann Theta Function}~\citep{riemann}, and it is easy to verify that $\Im(\mathbf{\Omega}) = \alpha\mL^T \mL$ is positive definite that guarantees convergence of Gaussian Lattice Sum.~\citet{riemann} further shows several upper bounds. In particular, given $R>0$,
\begin{align}
\label{eq:origingaussianbound}
\sum_{\vk \in \mathbb{Z}^d, \Vert \mL \vk + \vv\Vert \ge R} \Vert \mL \vk + \vv\Vert^p e^{-\Vert\mL\vk +\vv\Vert^2} \le \frac{d + p}{2} \left(\frac{2}{\rho}\right)^d \Gamma\left(\frac{d}{2}, \left(R -\frac{\rho}{2}\right)^2\right),
\end{align}
where $\Gamma(z,x)=\int_x^\infty t^{z-1}e^{-t}dt$ is the incomplete Gamma function and $\rho = \min \{\Vert \mL \vk \Vert \ |\ \vk\in \mathbb{Z}^d, \vk \ne\mathbf{0} \}$. Let $p = 0$ and a lattice matrix be $\sqrt{\pi \alpha} L$. We obtain 
\begin{align}
\label{eq:gaussianbound}
\sum_{\vk \in \mathbb{Z}^d, \sqrt{\pi \alpha } \Vert \mL \vk + \vv\Vert \ge R} e^{-\pi \alpha\Vert \mL \vk + \vv\Vert^2} \le \frac{d}{2} \left(\frac{2}{\rho}\right)^d \Gamma\left(\frac{d}{2}, \left(R -\frac{\rho}{2}\right)^2\right),
\end{align}
where $\rho = \min \{\sqrt{\pi\alpha}\Vert\mL \vk \Vert \ |\ \vk\in \mathbb{Z}^d, \vk \ne\mathbf{0} \}$. In addition, given $R = 0$, we can obtain the upper bound of the Gaussian Lattice Sum by incomplete Gamma function,
\begin{align}
\sum_{\vk \in \mathbb{Z}^d} e^{-\pi \alpha\Vert \mL \vk + \vv\Vert^2} \le \frac{d}{2} \left(\frac{2}{\rho}\right)^d \Gamma\left(\frac{d}{2}, \left(\frac{\rho}{2}\right)^2\right).
\end{align}


\section{Incomplete Bessel Function}

Let $x, y,\nu\in \mathbb{R}$, $x > 0$, $y\ge 0$, $\nu > 0$. The \textit{incomplete Bessel function}~\citep{bessel} has the below integral form
\begin{align}
\label{eq:bessel}
K_\nu(x,y) = \int_1^\infty t^{-\nu-1} e^{-xt - y/t} dt, 
\end{align}
In existing studies~\citep{bessel, besselcal, jones2007incomplete},
$K_\nu(x,y)$ is analytically continued to $\nu \in \mathbb{R}$. In this work, we follow these studies and also use the analytic continuation for our calculations, i.e., we also consider $\nu \in \mathbb{R}$.

\subsection{Fast Approximation of the Incomplete Bessel Function}
\label{fastbessel}
Computing the incomplete Bessel function is challenging as there does not exist direct closed-form solution. In this work, we investigate a fast approximation of the incomplete Bessel function.
Concretely, we adopt the algorithm in work by \citet{fft, besselcal}, where the incomplete Bessel function $K_\nu (x,y)$ is approximated by the $G_{n}^{(1)}$ transformation with the linear time complexity of $O(n)$, and $n$ here is the number of iterations. To begin with, the approximation $G_{n}^{(m)}$ to $\int_0^\infty f(t)dt$ is given as a solution of  
\begin{equation}
\label{eq:diff}
\frac{d^l}{dx^l}\left\{G_n^{(m)} - \int_0^x f(t)dt - \sum_{k=0}^{m-1}x^{\sigma_k}f^{(k)}(x) \sum_{i=0}^{n-1} \frac{\beta_{i,k}}{x^i}\right\} = 0,\quad l=0,1,\cdots, mn,
\end{equation}
where it is assumed that $\frac{d^l}{dx^l} G_n^{(m)} \equiv 0, \forall l > 0$, $\sigma_k = \min(s_k, k+1)$, and $s_k$ is the largest of $s\in \mathbb{Z}$ such that $\lim_{x\rightarrow \infty} x^s f^{(k)}(x) = 0$ for $k = 0,1,\cdots,m-1$. And we aim to obtain the unknown coefficients $\beta_{i,k}$ to approximate $\int_x^\infty f(t)dt$ from these $mn+1$ linear equations. By considering $l = 0$ in the Eqn.~(\ref{eq:diff}), we obtain
\begin{align}
G_n^{(1)} - \int_0^x f(t)dt = x^{\sigma_0}f(x) \sum_{i=0}^{n-1} \frac{\beta_{i,k}}{x^i}.
\end{align}
To eliminate the summation, \citet{besselcal} applied the $(x^2 \frac{d}{dx})$ operator $n$ times such that
\begin{align}
(x^2 \frac{d}{dx})^n \left[\frac{G_n^{(1)} - \int_0^x f(t)dt}{x^{\sigma_0 f(x)}} \right] = 0.
\end{align}
By doing this,
\begin{align}
G_n^{(1)} = \frac{(x^2 \frac{d}{dx} )^n (\frac{\int_0^x f(t)dt}{x^{\sigma_0}f(x)} )  }{ (x^2 \frac{d}{dx} )^n (\frac{1}{x^{\sigma_0}f(x)} ) } = \frac{\mathcal{N}_n(x)}{\mathcal{D}_n(x)},
\end{align}
where it sets
\begin{align}
\mathcal{N}_n(x) = (x^2\frac{d}{dx})\mathcal{N}_{n-1}(x) \quad \mbox{and} \quad \mathcal{D}_n(x) = (x^2\frac{d}{dx})\mathcal{G}_{n-1}(x),
\end{align}
and
\begin{align}
\mathcal{N}_0(x) = \frac{\int_0^x f(t)dt}{x^{\sigma_0} f(x)} \quad  \mbox{and} \quad \mathcal{D}_0(x) = \frac{1}{x^{\sigma_0} f(x)}.
\end{align}
This leads to a recursive algorithm for $G_n^{(1)}$ transformation. To compute the incomplete Bessel function $K_\nu(x,y)$, \citet{besselcal} investigated the following property
\begin{align}
K_\nu(x,y) + x^{\nu} \int_0^x  t^{-\nu-1} e^{-t - \frac{xy}{t}} dt = x^{\nu} \int_0^\infty  t^{-\nu-1} e^{-t - \frac{xy}{t}} dt,
\end{align}
where the term $\int_0^\infty  t^{-\nu-1} e^{-t - \frac{xy}{t}} dt$ can be approximated by $G_n^{(1)}$ transformation. Therefore, to approximate $K_\nu(x,y)$, we have corresponding approximation 
\begin{equation}
\begin{aligned}
\tilde{G}_n^{(1)} &= x^{\nu} (G_n^{(1)} -\int_0^x  t^{-\nu-1} e^{-t - \frac{xy}{t}} dt )\\
&= x^{\nu} \frac{ \mathcal{N}_n(x,y,\nu) - \mathcal{D}_n(x,y,\nu)\int_0^x  t^{-\nu-1} e^{-t - \frac{xy}{t}} dt }{\mathcal{D}_n(x,y,\nu)}\\
&= x^{\nu} \frac{\tilde{\mathcal{N}}_n(x,y,\nu)}{\mathcal{D}_n(x,y,\nu)}.
\end{aligned}
\end{equation}

As a result, the approximation $\tilde{G}_n^{(1)}$ to $K_\nu (x,y)$ is obtained by recursively solving $\tilde{\mathcal{N}}_n(x,y,\nu)$ and $\mathcal{D}_n(x,y,\nu)$~\citep{besselcal, gaudreau2012computation}. The detailed expressions of $\tilde{\mathcal{N}}_n(x)$ and $\mathcal{D}_n(x)$ are given in \citet{besselcal} Proposition 2.2. In addition, \citet{fft} further optimizes the approximation of the incomplete Bessel function when $\nu = 0$ and $x, y$ are both small. In this case, the remaining part of the Taylor expansion of $K_0(x,y)$ is small and we can approximate $K_0(x,y)$ by the first 
$m$ terms of the Taylor series such that
\begin{align}
K_0(x,y) \approx \sum_{n=0}^m \frac{(-1)^n}{n!} x^n y^n \Gamma(-n, x).
\end{align}
The detailed error bound by this expansion is shown in \citet{fft}.

\subsection{Convergence of Incomplete Bessel Function Summation}
\label{besselconverge}

Let $\nu,\alpha,\beta,\gamma \in \mathbb{R}$ and be constants. Given a $d$-dimensional lattice, we define the \textit{incomplete Bessel function summation} on the lattice as
\begin{align}
\sum_{\vk \in \mathbb{Z}^d, \pi\alpha \Vert \mL \vk + \vv \Vert^2 + \gamma > 0} K_\nu (\pi \alpha \Vert \mL \vk + \vv\Vert^2 + \gamma, \beta),
\end{align}
where $\alpha > 0$, $\beta \ge 0$, $\gamma \ge 0$, $\vv \in \mathbb{R}^d$ and $\mL \in \mathbb{R}^{d\times d}$ denoting the full rank lattice matrix. We aim to prove this summation of incomplete Bessel functions converges and can be approximated with an error bounded by the Gaussian Lattice Sum as introduced in Appendix~\ref{gls}. And this convergence property can be easily extended to the summation with $e^{2\pi i \zeta}$ coefficients since $| e^{2\pi i \zeta}K_\nu (x,y)|\le |K_\nu (x,y)| $.
\begin{proof} An upper bound of the incomplete Bessel function can be derived as
\begin{align}
\big|K_\nu(x,y)\big| &= \Big|\int_1^\infty t^{-\nu-1} e^{-xt - y/t} dt\Big|\le  \Big|\int_1^\infty t^{-\nu-1} e^{-xt} dt\Big| = \big|x^{-\nu} \Gamma(-\nu, x)\big|, 
\end{align}
where $\Gamma$ is the incomplete Gamma function described in Appendix~\ref{gls}. Based on this, we obtain
\begin{align}
\big|K_\nu (\pi \alpha \Vert \mL \vk + \vv\Vert^2 + \gamma, \beta)\big| \le \Big|\frac{\Gamma(-\nu, \pi \alpha \Vert \mL \vk + \vv\Vert^2 + \gamma)}{(\pi\alpha\Vert \mL \vk + \vv\Vert^2 + \gamma)^\nu}\Big|.
\end{align}
Given $x>0$, $|\frac{\Gamma(z, x)}{x^{z}}|$ has an upper bound~\citep{gamma} such that
\begin{equation}
\begin{aligned}
\Big|\frac{\Gamma(z, x)}{x^{z}}\Big|
&= \Big|e^{-x} \int_0^\infty e^{-x s} (1+s)^{z - 1} ds\Big| \\
&\le  \left \{
\begin{array}{ll}
    \frac{e^{-x}}{x - z + 1}, & \ z > 1\\
    \frac{e^{-x}}{x}, &  \ z \le 1
\end{array}.
\right.
\end{aligned}
\end{equation}
Considering a prefixed value $R\in \mathbb{R}$, $R^2 > -\nu$, $R^2 > 1$ and $R^2 > \gamma$, we have
\begin{equation}
\begin{aligned}
\epsilon(R) &=\sum_{\vk \in \mathbb{Z}^d, \pi\alpha| \mL \vk + \vv|^2 +\gamma \ge R^2 } K_\nu (\pi\alpha\Vert \mL \vk + \vv\Vert^2 +\gamma, \beta) \\
&\le \sum_{\vk \in \mathbb{Z}^d, \pi\alpha\Vert \mL \vk + \vv\Vert^2 +\gamma \ge R^2 } |K_\nu (\pi\alpha\Vert \mL \vk + \vv\Vert^2 +\gamma, \beta)| \\
&\le \sum_{\vk \in \mathbb{Z}^d,  \pi\alpha\Vert \mL \vk + \vv\Vert^2 +\gamma \ge R^2 } e^{-\pi\alpha\Vert \mL \vk + \vv\Vert^2 -\gamma} \\
&\le e^{-\gamma} E(\mL, \vv, \alpha)\\
&\le E(\mL, \vv, \alpha),
\end{aligned}
\end{equation}
where $E(\mL, \vv, \alpha)$ is the Gaussian Lattice Sum described in Appendix~\ref{gls}. Therefore, the incomplete Bessel function summation can be divided into two parts
\begin{align}
\sum_{\vk \in \mathbb{Z}^d, \pi\alpha \Vert \mL \vk + \vv \Vert^2 + \gamma > 0} K_\nu (\pi\alpha\Vert \mL \vk + \vv\Vert^2 +\gamma, \beta) = \sum_{\vk \in \mathbb{Z}^d, 0 < \pi\alpha\Vert \mL \vk + \vv\Vert^2 + \gamma \le R^2 } K_\nu (\pi\alpha\Vert \mL \vk + \vv\Vert^2 +\gamma, \beta) + \epsilon(R),
\end{align}
where the first part is a finite part inside an ellipsoid with a size of $\sqrt{R^2 /\pi \alpha - \gamma / \pi\alpha}$, and the second part $\epsilon(R)$ is bounded by the Gaussian Lattice Sum $E(\mL, \vv, \alpha)$ and is convergent. Therefore, the incomplete Bessel function summation is convergent. Consequently, to approximate the incomplete Bessel function summation, we can choose to evaluate the summation inside an ellipsoid with the size of $\sqrt{R^2 / \pi \alpha - \gamma /\pi \alpha}$ for a prefixed $R\in \mathbb{R}$, such that $R^2 > -\nu$, $R^2 > 1$ and $R^2 > \gamma$. Then the error $\epsilon(R)$ is bounded by Gaussian Lattice Sum $E(\mL, \vv, \alpha)$. We can further bound the error by the inequality~(\ref{eq:gaussianbound}) such that 
\begin{align}
\label{eq:error}
\epsilon(R) \le \sum_{\vk \in \mathbb{Z}^d, \sqrt{\pi\alpha}\Vert \mL \vk + \vv\Vert \ge \sqrt{R^2- \gamma} } e^{-\pi\alpha\Vert \mL \vk + \vv\Vert^2} \le \frac{d}{2} (\frac{2}{\rho})^d \Gamma(\frac{d}{2}, ( \sqrt{R^2- \gamma}-\frac{\rho}{2})^2),
\end{align}
where $\rho = \min \{\sqrt{\pi\alpha}\Vert \mL \vk \Vert \ |\ \vk\in \mathbb{Z}^d, \vk \ne\mathbf{0} \}$. This completes the proof.
\end{proof}

\section{Fast Algorithm of Potential Summation}

\subsection{Integral Transformation}
\label{proof_34_1}
In the main sections of our paper, it is much clear to use node positions to deliver our ideas. However, in the below sections, we are more interested in the vectors between nodes rather than the positions of those nodes. Given a full rank lattice matrix $\mL \in \mathbb{R}^{d\times d}$ and a vector $\vv\in \mathbb{R}^d$ between two atoms inside the unit cell, we denote $G(\mL, \vv)$ as the potential summation and $U(\mL, \vv)$ as the potential function. Here, if a vector $\vv^{\prime}$ is not inside the unit cell, we can apply a simple transformation~\citep{epstein} to $\vv^{\prime}$ by converting it into the fractional coordinate, reducing mod 1, and then converting it back, i.e., \begin{equation}
\vv = \mL (\mL^{-1}\vv^{\prime}\mod 1)
\end{equation} 
For a potential summation $S(\va, \vb)$ of a crystal with a lattice matrix $\mL$, we have $S(\va, \vb) = G(\mL, \vv_{\va\vb})$. Based on these notations, we aim to prove that the summation of the three introduced potentials can be transformed into an integral form as 
\begin{align}
\label{general}
G(\mL, \vv) = D \int_0^\infty t^{C-1}(-\delta(\vv, B) + \sum_{\vk \in \mathbb{Z}^d} e^{-A\pi\Vert\mL\vk+\vv\Vert^{2} t - \frac{B}{t}}) dt, 
\end{align}
where $A,B,C,D$ are constants derived from the corresponding specific potential forms and $\delta$ is the generalized delta function such that $\delta(\vv, B) = 1$ if and only if $\vv = \mathbf{0}$ and $B = 0$, otherwise $\delta(\vv, B) = 0$.
\begin{proof}
1). For the potentials in the form of $U(\mL,\vv) = 1/{\Vert \mL\vk+\vv\Vert }^{2p}$ and $\Vert \mL\vk+\vv\Vert\ne 0$, we apply the Mellin transform such that
\begin{equation}
\begin{aligned}
M\{U \}(\mL, \vv) &= \int_{0}^{\infty} t^{p-1} e^{-t\left\Vert\mL\vn+\vv\right\Vert^{2}} d t \\
&=\frac{\Gamma(p)}{\left\Vert\mL\vn+\vv\right\Vert^{2p}}.
\end{aligned}
\end{equation}
Consequently, we obtain
\begin{equation}
\label{eq:epstein}
U(\mL,\vv) = \frac{1}{\Vert\mL\vk+\vv\Vert^{2p}} =\frac{1}{\Gamma(p)} \int_{0}^{\infty} t^{p-1} e^{-t\left\Vert\mL\vn+\vv\right\Vert^{2}} d t.
\end{equation}
Next, by deriving the summation,
\begin{equation}
\begin{aligned}
G(\mL, \vv)=\sum_{\vk\in\mathbb{Z}^d,\Vert \mL\vk+\vv\Vert\ne 0} \frac{1}{\Vert \mL\vk+\vv\Vert^{2p}} &= \frac{1}{\Gamma(p)} \int_0^\infty t^{p-1} (-\delta(\vv) + \sum_{\vk\in\mathbb{Z}^d}e^{-\Vert\mL\vk+\vv\Vert^2 t}) dt.
\end{aligned}
\end{equation}
Apparently, we can obtain $A = 1/\pi$, $B = 0$, $C = p$ and $D = 1/\Gamma(p)$ for the summation of $U(\mL,\vv) = 1/{\Vert\mL\vk+\vv\Vert}^{2p}$. The same result is also given in~\citet{epstein}.

2). For the potentials in the form of $U(\mL,\vv) = e^{-\alpha \Vert\mL\vk + \vv\Vert}$, we consider the inverse Laplace transform on $e^{-\alpha \sqrt{s}}$ as shown by \citet{bateman1954tables}, such that
\begin{align}
\label{eq:inverse}
\mathcal{L}^{-1}\{e^{-\alpha \sqrt{s}}\} =  \frac{a}{2 \sqrt{\pi}}t^{-\frac{3}{2}} e^{-\frac{\alpha^2}{4 t}}.
\end{align}
Therefore, by applying the Laplace transform in Eqn.~(\ref{eq:inverse}) we derive the integral form of $e^{-\alpha \Vert\mL\vk + \vv\Vert}$: 
\begin{equation}
\begin{aligned}
U(\mL,\vv) &= e^{-\alpha \Vert\mL\vk + \vv\Vert} \\
&= \frac{a}{2 \sqrt{\pi}} \int_{0}^{\infty} t^{-\frac{3}{2}} e^{- t\Vert\mL\vk + \vv\Vert^{2} -\frac{\alpha^2}{4 t}} d t \\
&=\frac{\alpha}{2 \pi} \int_{0}^{\infty} t^{-\frac{3}{2}} e^{-\pi t\Vert\mL\vk + \vv\Vert^{2}-\frac{\alpha^{2}}{4 \pi t}} d t\left(t\leftarrow\pi t\right).
\end{aligned}
\end{equation}
Next, by deriving the summation,
\begin{equation}
\begin{aligned}
G(\mL,\vv)=\sum_{\vk\in\mathbb{Z}^d} e^{-\alpha \Vert \mL\vk + \vv \Vert} &=\frac{\alpha}{2 \pi} \int_{0}^{\infty} t^{-\frac{3}{2}}\sum_{\vk\in\mathbb{Z}^d} e^{-\pi t\Vert\mL\vk + \vv\Vert^{2}-\frac{\alpha^{2}}{4 \pi t}} d t.
\end{aligned}
\end{equation}
Apparently, we can obtain $A = 1$, $B = \frac{\alpha^2}{ 4\pi}$, $C = -\frac{1}{2}$ and $D = \frac{\alpha}{2\pi}$ for the summation of $U(\mL,\vv) = e^{-\alpha \Vert\mL\vk + \vv\Vert}$.

This completes the proof.
\end{proof}

\subsection{Calculating Integral Summation}
\label{integralsummation}

A short mathematical summary of this section is presented in \url{https://github.com/divelab/AIRS/blob/main/OpenMat/PotNet/summary.pdf}. As shown in Sec.~\ref{algorithm}, $G(\mL, \vv)$ can be written as the summations in the direct space and the Fourier space: 
\begin{equation}
\begin{aligned}
G(\mL, \vv)
&= D \int_0^\infty t^{C-1}(-\delta(\vv, B) + \sum_{\vk \in \mathbb{Z}^d} e^{-A\pi\Vert\mL\vk+\vv\Vert^{2} t - \frac{B}{t}}) dt\\
&= D \int_0^1 t^{C-1}(-\delta(\vv, B) + \sum_{\vk \in \mathbb{Z}^d} e^{-A\pi\Vert\mL\vk+\vv\Vert^{2} t - \frac{B}{t}}) dt + D \int_1^\infty t^{C-1}(-\delta(\vv, B) + \sum_{\vk \in \mathbb{Z}^d} e^{-A\pi\Vert\mL\vk+\vv\Vert^{2} t - \frac{B}{t}}) dt \\
&= G\textsubscript{Fourier}(\mL, \vv) +  G\textsubscript{direct}(\mL, \vv),
\end{aligned}
\end{equation}
where $G\textsubscript{Fourier}(\mL, \vv) = D \int_0^1 t^{C-1}(-\delta(\vv, B) + \sum_{\vk \in \mathbb{Z}^d} e^{-A\pi\Vert\mL\vk+\vv\Vert^{2} t - \frac{B}{t}}) dt$ denotes the summation in Fourier space, and 
$G\textsubscript{direct}(\mL, \vv) =  D \int_1^\infty t^{C-1}(-\delta(\vv, B) + \sum_{\vk \in \mathbb{Z}^d} e^{-A\pi\Vert\mL\vk+\vv\Vert^{2} t - \frac{B}{t}}) dt$ denotes the summation in direct space. Below, we prove that both $G\textsubscript{direct}(\mL, \vv)$ and $G\textsubscript{Fourier}(\mL, \vv)$ can be deduced into the incomplete Bessel function summation.
\begin{proof}
For $G\textsubscript{direct}(\mL, \vv)$, by deriving in direct space,
\begin{equation}
\begin{aligned}
\label{eqn:direct}
    G\textsubscript{direct}(\mL, \vv) &=  D \int_1^\infty t^{C-1}(-\delta(\vv, B) + \sum_{\vk \in \mathbb{Z}^d} e^{-A\pi\Vert\mL\vk+\vv\Vert^{2} t - \frac{B}{t}}) dt \\
    &= D \int_1^\infty t^{C-1}(-\delta(\vv, B) + e^{-A\pi \Vert\vv \Vert^2 t - \frac{B}{t}})dt + D\int_1^\infty t^{C-1}\sum_{\vk \in \mathbb{Z}^d, \vk \ne \mathbf{0}} e^{-A\pi\Vert\mL\vk+\vv\Vert^{2} t - \frac{B}{t}}dt \\
    &= D \int_1^\infty t^{C-1}(-\delta(\vv, B) + e^{-A\pi \Vert\vv \Vert^2 t - \frac{B}{t}})dt + D\sum_{\vk \in \mathbb{Z}^d, \vk\ne \mathbf{0}}\int_1^\infty t^{C-1} e^{-A\pi\Vert\mL\vk+\vv\Vert^{2} t - \frac{B}{t}}dt \\
    &= D \int_1^\infty t^{C-1}(-\delta(\vv, B) + e^{-A\pi \Vert\vv \Vert^2 t - \frac{B}{t}})dt + D\sum_{\vk \in \mathbb{Z}^d, \vk\ne \mathbf{0}} K_{-C}(A\pi \Vert \mL \vk + \vv \Vert^2, B).
\end{aligned}
\end{equation}
If $\vv = \mathbf{0}$ and $B = 0$, 
\begin{equation}
   G\textsubscript{direct}(\mL, \vv) = D\sum_{\vk \in \mathbb{Z}^d, \vk\ne \mathbf{0}} K_{-C}(A\pi \Vert \mL \vk\Vert^2, 0);
\end{equation}
And if $\vv = \mathbf{0}$ and $B\ne 0$, 
\begin{equation}
\begin{split}
    G\textsubscript{direct}(\mL, \vv) &=D \int_1^\infty t^{C-1} e^{-B/t}dt +  D\sum_{\vk \in \mathbb{Z}^d, \vk\ne \mathbf{0}} K_{-C}(A\pi \Vert \mL \vk \Vert^2, B) \\
    &=  DB^C(\Gamma(-C) - \Gamma(-C,B)) +  D\sum_{\vk \in \mathbb{Z}^d, \vk\ne \mathbf{0}} K_{-C}(A\pi \Vert \mL \vk \Vert^2, B);
\end{split}
\end{equation}
Otherwise,
\begin{equation}
   G\textsubscript{direct}(\mL, \vv) = D\sum_{\vk \in \mathbb{Z}^d} K_{-C}(A\pi \Vert \mL \vk + \vv \Vert^2, B).
\end{equation}
Overall,
\begin{equation}
\label{eqn:gdirect}
    G\textsubscript{direct}(\mL, \vv) = D\delta(\vv) B^C(\Gamma(-C) - \Gamma(-C,B)) +  D\sum_{\vk \in \mathbb{Z}^d, \Vert \mL \vk + \vv \Vert\ne 0} K_{-C}(A\pi \Vert \mL \vk + \vv \Vert^2, B).
\end{equation}
Here we obtain $G\textsubscript{direct}(\mL, \vv)$ as the incomplete Bessel function summation. Inspired by the Ewald summation~\citep{ewald, epstein}, we consider $G\textsubscript{Fourier}(\mL, \vv)$ on the reciprocal lattice using the Poisson summation~\citep{epstein}: 
\begin{equation}
\label{eq:possion}
\sum_{\vk \in \mathbb{Z}^{d}} e^{-2\pi i \vw \cdot \mL \vk -\pi t\left\Vert \mL\vk+\vv\right\Vert^{2}}=\frac{t^{-\frac{d}{2}} e^{2\pi i \vw \cdot \vv}}{\operatorname{det} \mL} \sum_{\vk \in \mathbb{Z}^{d}} e^{2 \pi i \mL^{\prime}\vk \cdot \vv -\frac{\pi}{t}\left\Vert\mL^{\prime}\vk + \vw\right\Vert^{2}},
\end{equation}
where $\vw\in \mathbb{R}^d$ and $\vw = \mathbf{0}$ in our case, and $\mL^{\prime} = \mL (\mL^T\mL)^{-1}$ is the lattice matrix of the reciprocal lattice. Therefore, we derive
\begin{equation}
\begin{split}
G\textsubscript{Fourier}(\mL, \vv)
&= D \int_0^1 t^{C-1} (-\delta(\vv, B) +  \sum_{\vk \in \mathbb{Z}^d} e^{-A\pi\Vert\mL \vk + \vv\Vert^2 t - \frac{B}{t}}) dt\\
&= -D \delta(\vv, B)\int_0^1 t^{C-1}dt +  D\int_0^1 t^{C-1}\sum_{\vk \in \mathbb{Z}^d} e^{-A\pi\Vert\mL \vk + \vv\Vert^2 t - \frac{B}{t}} dt\\
&= -D \delta(\vv, B)\int_0^1 t^{C-1}dt +  D\sum_{\vk \in \mathbb{Z}^d}\int_0^1 t^{C-1} e^{-A\pi\Vert\mL \vk + \vv\Vert^2 t - \frac{B}{t}} dt\\
&=-D \delta(\vv, B)\int_0^1 t^{C-1}dt + \frac{D}{A^C}\sum_{\vk \in \mathbb{Z}^d} \int_0^{A} t^{C-1} e^{-\pi \Vert\mL \vk + \vv\Vert^2 t - \frac{AB}{t}} dt(t \leftarrow \frac{t}{A}) \\
&= -D \delta(\vv, B)\int_0^1 t^{C-1}dt + \frac{1}{\operatorname{det} \mL} \frac{D}{A^C} \sum_{\vk \in \mathbb{Z}^d} \int_0^{A} t^{C-\frac{d}{2}-1} e^{2 \pi i \mL^{\prime}\vn \cdot \vv -\frac{\pi}{t}\left\Vert\mL^{\prime}\vk\right\Vert^{2} - \frac{AB}{t}} dt\mbox{(Eqn.~(\ref{eq:possion}))} \\
&=  -D \delta(\vv, B)\int_0^1 t^{C-1}dt + \frac{1}{\operatorname{det} \mL} \frac{D}{A^{\frac{d}{2}}} \sum_{\vk \in \mathbb{Z}^d} \int_0^{1} t^{C-\frac{d}{2} - 1} e^{2 \pi i \mL^{\prime}\vk \cdot \vv -\frac{\pi}{At}\left\Vert\mL^{\prime}\vk\right\Vert^{2} - \frac{B}{t}} dt(t\leftarrow At ) \\
&=  -D \delta(\vv, B)\int_0^1 t^{C-1}dt +  \frac{1}{\operatorname{det} \mL} \frac{D}{A^{\frac{d}{2}}} \sum_{\vk \in \mathbb{Z}^d} \int_1^{\infty} t^{\frac{d}{2}-C-1} e^{2 \pi i \mL^{\prime}\vn \cdot \vv -\frac{\pi t}{A}\left\Vert\mL^{\prime}\vk\right\Vert^{2} - Bt} dt (t\leftarrow \frac{1}{t} ) \\
&=  -D \delta(\vv, B)\int_0^1 t^{C-1}dt +  \frac{1}{\operatorname{det} \mL} \frac{D}{A^{\frac{d}{2}}} \int_1^{\infty} t^{\frac{d}{2}-C-1} \sum_{\vk \in \mathbb{Z}^d} e^{2 \pi i \mL^{\prime}\vn \cdot \vv -\frac{\pi t}{A}\left\Vert\mL^{\prime}\vk\right\Vert^{2} - Bt} dt.
\end{split}
\end{equation}
If $B = 0$, we obtain
\begin{equation}
\begin{split}
\label{eqn:analyticorigin}
& G\textsubscript{Fourier}(\mL, \vv) \\
&=  -D \delta(\vv)\int_0^1 t^{C-1}dt +  \frac{1}{\operatorname{det} \mL} \frac{D}{A^{\frac{d}{2}}} \int_1^{\infty} t^{\frac{d}{2}-C-1} (1 + \sum_{\vk \in \mathbb{Z}^d, \vk\ne\mathbf{0}} e^{2 \pi i \mL^{\prime}\vn \cdot \vv -\frac{\pi t}{A}\left\Vert\mL^{\prime}\vk\right\Vert^{2}}) dt \\
&= 
-D \delta(\vv)\int_0^1 t^{C-1}dt + \frac{1}{\operatorname{det} \mL} \frac{D}{A^{\frac{d}{2}}} \int_1^{\infty} t^{\frac{d}{2}-C-1} dt + \frac{1}{\operatorname{det} \mL} \frac{D}{A^{\frac{d}{2}}} \sum_{\vk \in \mathbb{Z}^d, \vk\ne\mathbf{0}} \int_1^{\infty} t^{\frac{d}{2}-C-1} e^{2 \pi i \mL^{\prime}\vn \cdot \vv -\frac{\pi t}{A}\left\Vert\mL^{\prime}\vk\right\Vert^{2}} dt \\
&= 
-D \delta(\vv)\int_0^1 t^{C-1}dt + \frac{1}{\operatorname{det} \mL} \frac{D}{A^{\frac{d}{2}}} \int_1^{\infty} t^{\frac{d}{2}-C-1} dt + \frac{1}{\operatorname{det} \mL} \frac{D}{A^{\frac{d}{2}}} \sum_{\vk \in \mathbb{Z}^d, \vk\ne\mathbf{0}} e^{2 \pi i \mL^{\prime}\vk \cdot \vv} K_{C - \frac{d}{2}} (\frac{\pi \left\Vert\mL^{\prime}\vk\right\Vert^{2}}{A}, 0).
\end{split}
\end{equation}
By applying analytic continuation to domain of $C$ as discussed in Appendix~\ref{analytic}, we obtain
\begin{equation}
\begin{split}
\label{eqn:analyticresult}
    G\textsubscript{Fourier}(\mL, \vv) = - \frac{D\delta(\vv)}{C} + \frac{1}{\operatorname{det} \mL} \frac{D}{(C-\frac{d}{2})A^{\frac{d}{2}}} + \frac{1}{\operatorname{det} \mL} \frac{D}{A^{\frac{d}{2}}} \sum_{\vk \in \mathbb{Z}^d, \vk\ne\mathbf{0}} e^{2 \pi i \mL^{\prime}\vk \cdot \vv} K_{C - \frac{d}{2}} (\frac{\pi \left\Vert\mL^{\prime}\vk\right\Vert^{2}}{A}, 0)
\end{split}
\end{equation}
with poles $C=0$ and $C=\frac{d}{2}$; Otherwise, if $B\ne 0$,
\begin{equation}
\begin{split}
G\textsubscript{Fourier}(\mL, \vv) &= \frac{1}{\operatorname{det} \mL} \frac{D}{A^{\frac{d}{2}}}\sum_{\vk \in \mathbb{Z}^d} e^{2 \pi i \mL^{\prime}\vk \cdot \vv} K_{C - \frac{d}{2}} (\frac{\pi \left\Vert\mL^{\prime}\vk\right\Vert^{2}}{A} + B, 0).
\end{split}
\end{equation}
Together, we obtain
\begin{multline}
\label{eqn:gfourier}
    G\textsubscript{Fourier}(\mL, \vv) =-\frac{D\delta(\vv, B)}{C} + \frac{\delta(B)}{\operatorname{det} \mL} \frac{D}{(C-\frac{d}{2})A^{\frac{d}{2}}} + \frac{1}{\operatorname{det} \mL} \frac{D}{A^{\frac{d}{2}}} \sum_{\vk \in \mathbb{Z}^d, \pi \left\Vert\mL^{\prime}\vk\right\Vert^{2}/A + B > 0} e^{2 \pi i \mL^{\prime}\vk \cdot \vv} K_{C - \frac{d}{2}} (\frac{\pi \left\Vert\mL^{\prime}\vk\right\Vert^{2}}{A} + B, 0).
\end{multline}
Apparently, $G\textsubscript{Fourier}(\mL, \vv)$ is deduced into the incomplete Bessel function summation, and this completes the proof.
\end{proof}
Therefore, both $G\textsubscript{Fourier}(\mL, \vv)$ and $G\textsubscript{direct}(\mL, \vv)$ can be expressed by the incomplete Bessel function summation. In addition, as shown in Appendix~\ref{besselconverge}, the incomplete Bessel function summation $\sum_{\vk \in \mathbb{Z}^d, \pi\alpha \Vert \mL \vk + \vv \Vert^2 + \gamma > 0} K_\nu(\pi\alpha\Vert\mL\vk + \vv\Vert^2 + \gamma, \beta)$ converges and can be approximated. Therefore, $G(\mL, \vv)$ converges and also can be approximated.

\subsection{Analytic Continuation of Potential Summations}
\label{analytic}

To represent the series that is not well-defined in its original domain, including the Coulomb potential summation $\sum_{\vk\in \mathbb{Z}^d,\Vert\mL\vk + \vv\Vert\ne0}\frac{1}{\Vert\mL\vk + \vv\Vert}$, we need to investigate the analytic continuation of the potential summation. Analytic continuation is a technique to extend the domain $P$ of a given analytic function $f(x)$. If there exists a domain $Q$ containing $P$, and a function $\hat{f}(x)$ that is analytic on $Q$, and $\hat{f}(x) = f(x)$ holding for all $x$ in $P$, consequently, $\hat{f}(x)$ is an analytic continuation of $f(x)$ to $Q$. As shown by \citet{kung2003complex}, the analytic continuation is unique and satisfies the permanence of functional relationships, i.e., the equations holding for $f(x)$ will also hold for $\hat{f}(x)$. 

In our case, we can expand the domain of $C$ in $G(\mL, \vv)$ to $C\in \mathbb{R}/\{0,\frac{d}{2}\}$ such that $G(\mL, \vv)$ is well-defined for any $C\in \mathbb{R}/\{0,\frac{d}{2}\}$. This is enabled by analytic continuation in Eqn.~(\ref{eqn:analyticorigin}). To be concrete, the original domain of $C$ is $(\frac{d}{2},+\infty)$ in Eqn.~(\ref{eqn:analyticorigin}) and we can extend the domain of $C$ for $\int_0^1 t^{C-1}dt$, $\int_1^\infty t^{\frac{d}{2}-C-1}dt$, and $K_{C-\frac{d}{2}}(\frac{\pi \left\Vert\mL^{\prime}\vk\right\Vert^{2}}{A}, 0)$, respectively. By applying the analytically continued incomplete Bessel function we can extend the domain of $C$ to $\mathbb{R}$. On the other hand, the analytic continuation on $\int_0^1 t^{C-1}dt$ and $\int_1^\infty t^{\frac{d}{2}-C-1}dt$ will result in two poles $C=0$ and $C=\frac{d}{2}$ on $\mathbb{R}$. Therefore, the final analytically continued domain of $C$ is $\mathbb{R}/\{0,\frac{d}{2}\}$. For the potential summation $\sum_{\vk \in \mathbb{Z}^d, \Vert\mL\vk + \vv\Vert\ne 0} 1/\Vert\mL\vk + \vv\Vert^{2p}$, we have $A=1/\pi, B=0, C = p, D=1/\Gamma(p)$ as shown in Appendix~\ref{proof_34_1}. By analytic continuation, we are able to compute the summation for any $p\in \mathbb{R}/\{\frac{d}{2},0,-1,-2,-3,\cdots\}$. Since $p=1$ and $d=3$ for our crystal dataset, we are able to compute the Coulomb potential summation. In next section, a more general analytic continuation result $p\in \mathbb{C}/\{ 0, \frac{d}{2}\}$ is given.

\subsection{Generalized Epstein Zeta Function and Analytic Continuation}
\label{sec:zeta}

In fact, the potential summation $\sum_{\vk \in \mathbb{Z}^d,\Vert\mL\vk+\vv\Vert\ne 0 }1/{\Vert\mL\vk+\vv\Vert}^{2p}$ is a special case of \textit{generalized Epstein zeta function}~\citep{crandall1987elementary, terras1973bessel, epstein}, and is a generalization of the Riemann zeta function. Let $s \in \mathbb{C}/\{0,d\}$, $\mL \in \mathbb{R}^{d\times d}$ and $\vv, \vu\in \mathbb{R}^d$ inside the unit cell and reciprocal unit cell respectively. The generalized Epstein zeta function~\citep{epstein} has the below summation form
\begin{equation}
Z_{\mL}(s; \vu, \vv) = \sum_{\vk \in \mathbb{Z}^d, \Vert\mL \vk - \vv \Vert\ne 0} \frac{e^{2\pi i \vu \cdot \mL \vk}}{\left\Vert\mL \vk - \vv \right\Vert^s}.
\end{equation}
Apparently, the potential summation $\sum_{\vk \in \mathbb{Z}^d,\Vert\mL\vk+\vv\Vert\ne 0 }1/{\Vert\mL\vk+\vv\Vert}^{2p}$ can be expressed in terms of generalized Epstein zeta function as $Z_{\mL}(2p; \mathbf{0}, -\vv)$. In addition, $Z_{\mL}(s; \vu, \vv)$ has an analytic continuation to the entire complex plane, except for simple poles at $s = 0$ and $s = d$~\citep{crandall1987elementary}, which is corresponding to our result in Appendix~\ref{analytic}. Moreover, $Z_{\mL}(s; \vu, \vv)$ can be written in the form of an integral summation~\citep{epstein} similar to Eqn.~(\ref{eq:epstein}) such that
\begin{equation}
Z_{\mL}(s; \vu, \vv) = \sum_{\vk \in \mathbb{Z}^d, \Vert\mL \vk - \vv \Vert\ne 0} \frac{e^{2\pi i \vu \cdot \mL \vk}}{\left\Vert\mL \vk - \vv \right\Vert^s} = \frac{\pi^{s/2} }{\Gamma(s/2)} \int_{0}^{\infty} t^{s/2-1} (-\delta(\vv) + \sum_{\vk \in \mathbb{Z}^d}e^{2\pi i \vu \cdot \mL \vk-\pi t\left\Vert \mL\vk-\vv\right\Vert^{2}}) d t.
\end{equation}
Based on this, we can also split the integral and apply Poisson summation in Eqn.~(\ref{eq:possion}) to obtain two summations of incomplete Bessel functions to evaluate this series. To be concrete, similar to Eqn.~(\ref{eqn:direct}), the summation of generalized Epstein zeta function on direct space is
\begin{equation}
\begin{split}
    G\textsubscript{direct}(s;\mL,\vu ,\vv) &= \frac{\pi^{s/2} }{\Gamma(s/2)} \int_{1}^{\infty} t^{s/2-1} (-\delta(\vv) + \sum_{\vk \in \mathbb{Z}^d}e^{2\pi i \vu \cdot \mL \vk-\pi t\left\Vert \mL\vk-\vv\right\Vert^{2}}) d t\\
    &= \frac{\pi^{s/2} }{\Gamma(s/2)} \int_{1}^{\infty} t^{s/2-1} (-\delta(\vv) + e^{-\pi t\left\Vert\vv\right\Vert^{2}})dt + \sum_{\vk \in \mathbb{Z}^d, \vk\ne \mathbf{0}}\frac{\pi^{s/2} }{\Gamma(s/2)} \int_{1}^{\infty} t^{s/2-1}e^{2\pi i \vu \cdot \mL \vk-\pi t\left\Vert \mL\vk-\vv\right\Vert^{2}} d t \\
    &= \frac{\pi^{s/2} }{\Gamma(s/2)} \int_{1}^{\infty} t^{s/2-1} (-\delta(\vv) + e^{-\pi t\left\Vert\vv\right\Vert^{2}})dt + \frac{\pi^{s/2} }{\Gamma(s/2)} \sum_{\vk \in \mathbb{Z}^d, \vk\ne \mathbf{0}}e^{2\pi i \vu \cdot \mL \vk} K_{-s/2}(\pi \Vert \mL \vk - \vv\Vert^2, 0).
\end{split}
\end{equation}
By further exploring two cases of $\delta(\vv)$, we obtain
\begin{equation}
    G\textsubscript{direct}(s;\mL,\vu ,\vv) = \frac{\pi^{s/2} }{\Gamma(s/2)} \sum_{\vk \in \mathbb{Z}^d, \Vert\mL\vk -\vv\Vert \ne 0}e^{2\pi i \vu \cdot \mL \vk} K_{-s/2}(\pi\Vert \mL \vk - \vv\Vert^2, 0).
\end{equation}
The summation of generalized Epstein zeta function on Fourier space is
\begin{equation}
\begin{split}
    G\textsubscript{Fourier}(s;\mL,\vu ,\vv) &= \frac{\pi^{s/2} }{\Gamma(s/2)} \int_{0}^{1} t^{s/2-1} (-\delta(\vv) + \sum_{\vk \in \mathbb{Z}^d}e^{2\pi i \vu \cdot \mL \vk-\pi t\left\Vert \mL\vk-\vv\right\Vert^{2}}) d t \\
    &= -\frac{\delta(\vv)\pi^{s/2} }{\Gamma(s/2)} \int_{0}^{1} t^{s/2-1}dt + \frac{\pi^{s/2} }{\Gamma(s/2)}\int_{0}^{1} t^{s/2-1}\sum_{\vk \in \mathbb{Z}^d}e^{2\pi i \vu \cdot \mL \vk-\pi t\left\Vert \mL\vk-\vv\right\Vert^{2}}d t \\
    &=\!\begin{multlined}[t]
        -\frac{\delta(\vv)\pi^{s/2} }{\Gamma(s/2)} \int_{0}^{1} t^{s/2-1}dt +\\ \frac{\pi^{s/2}e^{2\pi i \vu \cdot \vv} }{\Gamma(s/2)\operatorname{det} \mL}\int_{0}^{1} t^{s/2-d/2-1}\sum_{\vk \in \mathbb{Z}^d}e^{-2\pi i \vv \cdot \mL^{\prime} \vk-\frac{\pi}{t}\left\Vert \mL^{\prime}\vk-\vu\right\Vert^{2}} d t (Eqn.~(\ref{eq:possion})) 
    \end{multlined} \\
     &= \!\begin{multlined}[t]
        -\frac{\delta(\vv)\pi^{s/2} }{\Gamma(s/2)} \int_{0}^{1} t^{s/2-1}dt +\\ \frac{\pi^{s/2}e^{2\pi i \vu \cdot \vv} }{\Gamma(s/2)\operatorname{det} \mL}\int_{1}^{\infty} t^{d/2-s/2-1}\sum_{\vk \in \mathbb{Z}^d}e^{-2\pi i \vv \cdot \mL^{\prime} \vk-\pi t\left\Vert \mL^{\prime}\vk-\vu\right\Vert^{2}} d t (t\leftarrow \frac{1}{ t}).
    \end{multlined}
\end{split}
\end{equation}
if $\vu = \mathbf{0}$,
\begin{equation}
\begin{split}
    G\textsubscript{Fourier}(s;\mL,\vu ,\vv) &= \!\begin{multlined}[t]
        -\frac{\delta(\vv)\pi^{s/2} }{\Gamma(s/2)} \int_{0}^{1} t^{s/2-1}dt +\\ \frac{\pi^{s/2}}{\Gamma(s/2)\operatorname{det} \mL}\int_{1}^{\infty} t^{d/2-s/2-1}\sum_{\vk \in \mathbb{Z}^d}e^{-2\pi i \vv \cdot \mL^{\prime} \vk-\pi t\left\Vert \mL^{\prime}\vk\right\Vert^{2}} d t
    \end{multlined} \\
    &= 
    \!\begin{multlined}[t]
        -\frac{\delta(\vv)\pi^{s/2} }{\Gamma(s/2)} \int_{0}^{1} t^{s/2-1}dt +\\ \frac{\pi^{s/2} }{\Gamma(s/2)\operatorname{det} \mL}\int_{1}^{\infty} t^{d/2-s/2-1}(1 + \sum_{\vk \in \mathbb{Z}^d, \vk\ne \mathbf{0}}e^{-2\pi i \vv \cdot \mL^{\prime} \vk-\pi t\left\Vert \mL^{\prime}\vk\right\Vert^{2}}) d t
    \end{multlined} \\
    &= 
    \!\begin{multlined}[t]
        -\frac{\delta(\vv)\pi^{s/2} }{\Gamma(s/2)} \int_{0}^{1} t^{s/2-1}dt + \frac{\pi^{s/2} }{\Gamma(s/2)\operatorname{det} \mL}\int_{1}^{\infty} t^{d/2-s/2-1}dt + \\
        \frac{\pi^{s/2}}{\Gamma(s/2)\operatorname{det} \mL}\sum_{\vk \in \mathbb{Z}^d, \vk\ne \mathbf{0}}\int_{1}^{\infty} t^{d/2-s/2-1}e^{-2\pi i \vv \cdot \mL^{\prime} \vk-\pi t\left\Vert \mL^{\prime}\vk\right\Vert^{2}} d t
    \end{multlined} \\
    &= 
    \!\begin{multlined}[t]
        -\frac{\delta(\vv)\pi^{s/2} }{\Gamma(s/2)} \int_{0}^{1} t^{s/2-1}dt + \frac{\pi^{s/2} }{\Gamma(s/2)\operatorname{det} \mL}\int_{1}^{\infty} t^{d/2-s/2-1}dt + \\
        \frac{\pi^{s/2}}{\Gamma(s/2)\operatorname{det} \mL}\sum_{\vk \in \mathbb{Z}^d, \vk\ne \mathbf{0}}e^{-2\pi i \vv \cdot \mL^{\prime} \vk} K_{s/2-d/2}(\pi \Vert \mL^{\prime}\vk\Vert, 0);
    \end{multlined}
\end{split}
\end{equation}
Otherwise,
\begin{equation}
\begin{split}
    G\textsubscript{Fourier}(s;\mL,\vu ,\vv) &= \!\begin{multlined}[t]
        -\frac{\delta(\vv)\pi^{s/2} }{\Gamma(s/2)} \int_{0}^{1} t^{s/2-1}dt + \\
        \frac{\pi^{s/2}e^{2\pi i \vu \cdot \vv}}{\Gamma(s/2)\operatorname{det} \mL}\sum_{\vk \in \mathbb{Z}^d}e^{-2\pi i \vv \cdot \mL^{\prime} \vk} K_{s/2-d/2}(\pi \Vert \mL^{\prime}\vk -\vu \Vert, 0).
    \end{multlined}
\end{split}
\end{equation}
Then
\begin{equation}
\begin{split}
    G\textsubscript{Fourier}(s;\mL,\vu ,\vv) &= 
    \!\begin{multlined}[t]
        -\frac{\delta(\vv)\pi^{s/2} }{\Gamma(s/2)} \int_{0}^{1} t^{s/2-1}dt + \frac{\delta(\vu)\pi^{s/2}  }{\Gamma(s/2)\operatorname{det} \mL}\int_{1}^{\infty} t^{d/2-s/2-1}dt + \\
        \frac{\pi^{s/2}e^{2\pi i \vu \cdot \vv}}{\Gamma(s/2)\operatorname{det} \mL}\sum_{\vk \in \mathbb{Z}^d, \vk\ne \mathbf{0}}e^{-2\pi i \vv \cdot \mL^{\prime} \vk} K_{s/2 - d/2}(\pi \Vert \mL^{\prime}\vk -\vu\Vert, 0);
    \end{multlined}
\end{split}
\end{equation}
Therefore,
\begin{equation}
\begin{split}
    Z_\mL(s;\vu,\vv) &= G\textsubscript{direct}(s;\mL,\vu ,\vv) + G\textsubscript{Fourier}(s;\mL,\vu ,\vv) \\
    &= \!\begin{multlined}[t]
        \frac{\pi^{s/2} }{\Gamma(s/2)} \sum_{\vk \in \mathbb{Z}^d, \Vert\mL\vk -\vv\Vert \ne 0}e^{2\pi i \vu \cdot \mL \vk} K_{-s/2}(\pi\Vert \mL \vk - \vv\Vert^2, 0)  -\frac{\delta(\vv)\pi^{s/2} }{\Gamma(s/2)} \int_{0}^{1} t^{s/2-1}dt + \\ \frac{\delta(\vu)\pi^{s/2}  }{\Gamma(s/2)\operatorname{det} \mL}\int_{1}^{\infty} t^{d/2-s/2-1}dt +
        \frac{\pi^{s/2}e^{2\pi i \vu \cdot \vv}}{\Gamma(s/2)\operatorname{det} \mL}\sum_{\vk \in \mathbb{Z}^d, \vk\ne \mathbf{0}}e^{-2\pi i \vv \cdot \mL^{\prime} \vk} K_{s/2 - d/2}(\pi \Vert \mL^{\prime}\vk -\vu\Vert, 0).
    \end{multlined}
\end{split}
\end{equation}
Similar to Eqn.~(\ref{eqn:analyticorigin}), we apply analytic continuation and obtain
\begin{equation}
\begin{split}
\label{eqn:generalepstein}
    Z_\mL(s;\vu,\vv) &= G\textsubscript{direct}(s;\mL,\vu ,\vv) + G\textsubscript{Fourier}(s;\mL,\vu ,\vv) \\
    &= \!\begin{multlined}[t]
        \frac{\pi^{s/2} }{\Gamma(s/2)} \sum_{\vk \in \mathbb{Z}^d, \Vert\mL\vk -\vv\Vert \ne 0}e^{2\pi i \vu \cdot \mL \vk} K_{-s/2}(\pi\Vert \mL \vk - \vv\Vert^2, 0)  -\frac{\delta(\vv)\pi^{s/2} }{\Gamma(s/2)s/2} + \\ \frac{\delta(\vu)\pi^{s/2}  }{\Gamma(s/2)\operatorname{det} \mL (d/2 - s/2)} +
        \frac{\pi^{s/2}e^{2\pi i \vu \cdot \vv}}{\Gamma(s/2)\operatorname{det} \mL}\sum_{\vk \in \mathbb{Z}^d, \vk\ne \mathbf{0}}e^{-2\pi i \vv \cdot \mL^{\prime} \vk} K_{s/2 - d/2}(\pi \Vert \mL^{\prime}\vk -\vu\Vert, 0).
    \end{multlined}
\end{split}
\end{equation}
Here,~\citet{epstein} gives the same result Eqn.~(\ref{eqn:generalepstein}) with incomplete Gamma function, while we provide detailed derivation on $Z_{\mL}(s; \vu, \vv)$. For more details of the generalized Epstein zeta function, we refer readers to~\citet{crandall1987elementary, terras1973bessel, epstein, kirsten1994generalized, selberg1967epstein}. The generalized Epstein zeta function is used for the computation of Madelung constants~\citep{epstein} as described in Appendix~\ref{born-lande}. This is because one can view the term $e^{2\pi i \vu \cdot \mL \vk} = \cos (2\pi \vu \cdot \mL \vk) + i \sin (2\pi \vu \cdot \mL \vk)$ as the charge distribution in the crystal system. If the phase $2\vu \cdot \mL \vk \in \mathbb{Z}$ for any $\vk$, the generalized Epstein zeta function will become an alternating series. This is useful for the ionic crystal systems where the unit cell is generally neutral and the Coulomb potentials cancel each other. A famous example is the Madelung constant of $NaCl$, which is derived by the summation of Coulomb potentials among $Na$ and $Cl$ ions. By using the generalized Epstein zeta function, we have the Madelung constant of $NaCl$ as $M=Z_{1_3}(1;(\frac{1}{2})_3,0_3)$~\citep{epstein} and here $a_n$ denotes a diagonal matrix with all main diagonal values as the scalar $a$. 

One can also calculate the Madelung constant by considering extracting terms in $Z_\mL(s;\vu,\vv)$ with the same coefficient $e^{2\pi i \vu \cdot \mL \vk}$ as individual potential summations and computing those individual summations (by analytic continuation). This is due to the fact that $Z_\mL(s;\vu,\vv)$ and these individual summations are all calculated by incomplete Bessel functions and share the same analytically continued domain. In other words, $Z_\mL(s;\vu,\vv)$ can be calculated by a linear combination of individual potential summations (by analytic continuation). It is useful for the case where $\vu$ is initially unknown but can be learned. Here we give a simple numerical example to calculate eta function $\sum_{n=1}^{\infty} \frac{(-1)^{n+1}}{\sqrt{n}}$ by analytic continuation such that 
\begin{equation}
\sum_{n=1}^\infty \frac{(-1)^{n+1}}{\sqrt{n}} =  \sum_{n=0}^\infty \frac{(-1)^n}{\sqrt{n + 1}}
= \frac{1}{\sqrt{2}}\sum_{n=0}^\infty (\frac{1}{\sqrt{n + \frac{1}{2}}} -  \frac{1}{\sqrt{n + 1}})
= \frac{1}{\sqrt{2}}\left(\zeta(\frac{1}{2}, \frac{1}{2})-\zeta(\frac{1}{2}, 1)\right),
\end{equation}
where $\zeta(s, a)$ is Hurwitz zeta function and $\zeta(s, a) = \sum_{n=0}^\infty \frac{1}{(n+a)^s}$ when $s > 1, a\ne 0,-1,-2,...$, and its analytic continuation elsewhere. That is, we can use the analytically continued zeta function $\zeta(s,a)$ to precisely evaluate a conditionally convergent series (or alternating zeta function). Overall, the above shows that we can use analytically continued potential summations to approximate the total contribution of potentials where individual potential summations cancel each other. We further show the Madelung constant calculation of $NaCl$ by analytic continuation in Appendix~\ref{born-lande}.  

\subsection{Implementation and Numerical Examples of Approximation}
\label{example}

\begin{table}[t]
\caption{Numerical examples of our algorithm. We approximate results by using all grid elements and approximating the error upper bound. The first column denotes the ground truth targets we aim to approximate. By choosing different grid length $2 * r$ in our algorithm, we can obtain different evaluation results and different estimated error upper bound of the approximation as described in the second, the third, and the fourth column. The fifth column gives the truth error between our approximation and ground truth. And the sixth column denotes time cost of our approximation. Here, $\zeta(x) = \sum_{n=1}^\infty 1/n^x$. Implementation details can be found in Appendix~\ref{example}.}
\label{table:algorithm}
\begin{center}
\begin{tabular}{lccccc}
\toprule
Ground Truth  & Evaluation & $r$ & Estimated Error & True Error & Time \\ 
\midrule
$2*\zeta (2)=3.28986813$ & 3.28068288 & 1 & 9e-1 & 9e-3 & 2 ms \\
$2*\zeta (2)=3.28986813$ & 3.28984070 & 2 & 6e-2 & 3e-5 & 3 ms \\
$2*\zeta (2)=3.28986813$ & 3.28986812 & 3 & 7e-4 & 1e-8 & 3 ms \\
$2*\zeta (2)=3.28986813$ & 3.28986813 & 4 & 1e-6 & $<$ 1e-8 & 3 ms \\
$2*\zeta (3)=2.40411381$ & 2.40411381 & 4 & 1e-6 & $<$ 1e-8 & 3 ms \\
$2*\zeta (4)=2.16464647$ & 2.16464647 & 4 & 1e-6 & $<$ 1e-8 & 3 ms \\
$\sum_{\vn\in \mathbb{Z}^2, \vn\ne \mathbf{0}} \frac{1}{|\vn|^4} = 6.02681204$ & 5.99068949 & 1 & 3 & 4e-2 & 3 ms\\
$\sum_{\vn\in \mathbb{Z}^2, \vn\ne \mathbf{0}} \frac{1}{|\vn|^4} = 6.02681204$ & 6.02670959 & 2 & 4e-1 & 1e-4 & 3 ms\\
$\sum_{\vn\in \mathbb{Z}^2, \vn\ne \mathbf{0}} \frac{1}{|\vn|^4} = 6.02681204$ & 6.02681199 & 3 & 7e-3 & 5e-8 & 3 ms\\
$\sum_{\vn\in \mathbb{Z}^2, \vn\ne \mathbf{0}} \frac{1}{|\vn|^4} = 6.02681204$ & 6.02681204 & 4 & 2e-5 & $<$ 1e-8  & 3 ms\\
$\sum_{n\in \mathbb{Z}} e^{-|n |} = 2.16395341$ & 2.16395326 & 1 & 4e-1 & 2e-7 & 2 ms\\
$\sum_{n\in \mathbb{Z}} e^{-|n |} = 2.16395341$ & 2.16395341 & 2 & 3e-4 & $<$ 1e-8 & 4 ms\\
$\sum_{\vn\in \mathbb{Z}^3} e^{-| \vn |} = 25.39268269$ & 25.39268214 & 1 & 2.5 & 5e-7 & 3 ms\\
$\sum_{\vn\in \mathbb{Z}^3} e^{-| \vn |} = 25.39268269$ & 25.39268269 & 2 & 1e-2 & $<$ 1e-8 & 3 ms \\
\bottomrule
\end{tabular}
\end{center}
\end{table}

In this section, we explicate the methodology applied for the implementation of our algorithm. A predefined $d$-dimensional discrete grid, centered at the origin and having a length defined by $2 r$, is applied for point selection of summation approximation. Typically, $r> R$ is delineated in the algorithm. Owing to the complexities associated with computing the inverse of the incomplete Gamma function, our primary objective is to establish the value of $R$, after which we calculate its corresponding error bound. To elaborate, we commence by determining the value of $R$ within a crystalline structure, following which we ascertain the associated error bound, as prescribed by Eqn.~(\ref{eq:error}). Within the $d$-dimensional discrete grid, a set of points inside an ellipsoid predicated on $R$ is selected for the evaluation of the summation of incomplete Bessel functions. It is essential to underscore that determining $R$ for each crystal is a nontrivial procedure. In cases where $R$ is not predetermined, we employ all grid elements to approximate the result and estimate the error upper bound by approximating $R$. It is also worth noting that we can take advantage of operations with vectorization when summing over grid elements and thus boost the final computation speed. As $\vv$ is inside the lattice, the approximation of $R$ is achieved by considering $r_{min}$, which is the length of the minor axis of an ellipsoid. And the ellipsoid is constructed via the expression $\mL\vx + \vv$ or $\mL' \vx$, with $\vx$ denoting points derived from a sphere centered at the origin with a radius $r$. Formally, given a lattice matrix $\mL\in \mathbb{R}^{d\times d}$, a vector $\vv \in \mathbb{R}^d$ inside a unit cell, and the constants $A,B,C,D$ derived from specific potential functions as described in Appendix~\ref{proof_34_1}, we aim to evaluate these two parts $G\textsubscript{direct}(\mL,\vv)$ and $G\textsubscript{Fourier}(\mL,\vv)$.

To evaluate $G\textsubscript{direct}(\mL,\vv)$ according to Eqn.~(\ref{eqn:gdirect}), we derive the following steps. 

\emph{\textbf{Step 1:}} Determine the grid residing in the $d$-dimensional integer space $\mathbb{Z}^d$ with length denoted as $2r$ and a value $R$ such that the conditions $r > R$ and $R^2 \ge C$, $R^2 \ge 1$ are satisfied. Subsequently, select points denoted by $\vk$ from the grid, adhering to the inequality $A\pi\Vert \mL \vk + \vv\Vert^2 \le R^2$. Once the points are selected, calculate the function represented by Eqn.~(\ref{eqn:gdirect}). In the scenario where $R$ remains undefined, it is recommended to use large $r$ and select all available points on the grid for computation of Eqn.~(\ref{eqn:gdirect}). $R$ is then approximated by $\sqrt{A\pi}(r_{min} - \Vert \vv\Vert)$.

\emph{\textbf{Step 2:}} 
Compute the error bound denoted as $\epsilon$ by the following formula: $\epsilon = \frac{d}{2} (\frac{2}{\rho})^d \Gamma(\frac{d}{2}, (R - \frac{\rho}{2})^2)$, where $\rho$ satisfies the condition $\rho = \min\{\sqrt{A\pi}\Vert\mL \vk\Vert \ |\ \vk\ne\mathbf{0} \}$.

To evaluate $G\textsubscript{Fourier}(\mL,\vv)$ according to Eqn.~(\ref{eqn:gfourier}), we derive the following steps.

\emph{\textbf{Step 1:}} Determine the grid residing in the $d$-dimensional integer space $\mathbb{Z}^d$ with length denoted as $2r$ and a value $R$ such that the conditions $r > R$ and $R^2 \ge \frac{d}{2} - C$, $R^2 \ge B$, $R^2 \ge 1$ are satisfied. Subsequently, select points denoted by $\vk$ from the grid, adhering to the inequality $\frac{\pi}{A}\Vert \mL' \vk\Vert^2 + B \le R^2$. Once the points are selected, calculate the function represented by Eqn.~(\ref{eqn:gfourier}). In the scenario where $R$ remains undefined, it is recommended to use large $r$ and select all available points on the grid for computation of Eqn.~(\ref{eqn:gfourier}). $R$ is then approximated by $\sqrt{\frac{\pi}{A}r_{min}^2 + B}$.

\emph{\textbf{Step 2:}} Compute the error bound denoted as $\epsilon$ by the following formula: $\epsilon = \frac{d}{2} (\frac{2}{\rho})^d \Gamma(\frac{d}{2}, (\sqrt{R^2 - B} - \frac{\rho}{2})^2)$, where $\rho$ satisfies the condition $\rho = \min\{\sqrt{\frac{\pi}{A}} \Vert \mL^{\prime} \vk\Vert \ |\ \vk\in\mathbb{Z}^d, \vk\ne\mathbf{0} \}$.

Our implementation is based on Cython, GNU Scientific Library~\citep{galassi2002gnu} and ScaFaCoS~\citep{scafacos-web}, in which the native incomplete Gamma function and incomplete Bessel function are used. We show the evaluation examples in Table~\ref{table:algorithm} with the corresponding error bound and evaluation time. The running time is at the scale of milliseconds.

\subsection{Potential Summation Extensions}
\label{addition}

To highlight the versatility of our potential summation method, we incorporate additional potentials that can be computed using our algorithm and Laplace transform. These include the Lennard-Jones potential, Morse potential, and screened Coulomb potential, which typically find application in the analysis of specific categories of materials, thereby demonstrating the broad applicability of our method.

\textbf{Lennard-Jones Potential}~\citep{lj} is an intermolecular pair potential that is usually used for gas or organic materials. Let $\epsilon$ and $\sigma$ be hyperparameters. The commonly used expression for the Lennard-Jones potential is
\begin{equation}
U\textsubscript{LJ}(\mL, \vv) = 4\epsilon  \left[\left(\frac{\sigma}{\Vert\mL \vk + \vv\Vert}\right)^{12} - \left(\frac{\sigma}{\Vert\mL \vk + \vv\Vert}\right)^6 \right].
\end{equation}
And the summation of $U\textsubscript{LJ}(\mL, \vv)$ can be converted to two potential summations of $1/\Vert\mL\vk + \vv\Vert^{2p}$ with $p=3$ and $p=6$ respectively, such that
\begin{equation}
G\textsubscript{LJ}(\mL, \vv) = 4\epsilon \left[\sigma^{12}\left(\sum_{\vk\in\mathbb{Z}^d, \Vert\mL\vk + \vv\Vert\ne 0}\frac{1}{\Vert\mL\vk + \vv\Vert^{12}}\right)- \sigma^{6}\left(\sum_{\vk\in\mathbb{Z}^d, \Vert\mL\vk + \vv\Vert\ne 0}\frac{1}{\Vert\mL\vk + \vv\Vert^{6}}\right) \right],
\end{equation}
where we already show the calculation of the potential summation of $1/\Vert\mL\vk + \vv\Vert^{2p}$ in Appendix~\ref{proof_34_1}.

\textbf{Morse Potential}~\citep{morse1929diatomic} is an interatomic potential of diatomic molecules and is used for simple molecular materials. Let $D_e$ and $r_e$ be hyperparameters. The Morse potential has a mathematical form of
\begin{equation}
U\textsubscript{Morse}(\mL, \vv) = D_e \left(e^{-2a (\Vert\mL \vk + \vv\Vert - r_e)} - 2e^{-a (\Vert\mL \vk + \vv\Vert - r_e)}\right).
\end{equation}
Similarly, the summation of $U\textsubscript{Morse}(\mL, \vv)$ can be converted to two potential summations of $e^{-\alpha\Vert\mL\vk + \vv\Vert}$ with $\alpha = a$ and $\alpha = 2a$ respectively, such that
\begin{equation}
G\textsubscript{Morse}(\mL, \vv) = D_e \left(e^{2ar_e}\sum_{\vk\in\mathbb{Z}^d} e^{-2a \Vert\mL \vk + \vv\Vert} - 2 e^{ar_e}\sum_{\vk\in\mathbb{Z}^d}e^{-a\Vert\mL \vk + \vv\Vert}\right),
\end{equation}
where we already show the calculation of the potential summation of $e^{-\alpha \Vert\mL \vk + \vv\Vert}$ in Appendix~\ref{proof_34_1}.

\textbf{Screened Coulomb Potential} represents the Coulomb interactions with damping of electric fields. It is an important potential reflecting the behaviors of charge-carrying fluids or particles in semiconductors~\citep{kirichenko2021influence}. Let $e_{0}$ be elementary charge constant and $\alpha$ be a scaling hyperparameter. The screened Coulomb potential has an analytic form of $V(\va,\vb) = \frac{z_{\va}z_{\vb}e^2_{0}}{d(\va,\vb)}e^{-\alpha d(\va,\vb)}$, where $d(\va,\vb)$ is the distance between atom $\va$ and $\vb$, and $z_{\va},z_{\vb}$ are charges of atom $\va$ and $\vb$. Similar to our Coulomb potential case, since $z_{\va},z_{\vb},e_{0}$ are constants and can be extracted outside the summation, we derive a simplified screened Coulomb potential
\begin{equation}
\label{screened}
U\textsubscript{screened}(\mL,\vv) = \frac{e^{-\alpha \Vert\mL \vk + \vv\Vert}}{\Vert\mL \vk + \vv\Vert}.
\end{equation}
Consider the inverse Laplace transform on $e^{-\alpha \sqrt{s}}/\sqrt{s}$~\citep{bateman1954tables}, we obtain
\begin{equation}
\mathcal{L}^{-1}\{e^{-\alpha \sqrt{s}}/\sqrt{s}\} = \frac{1}{\sqrt{\pi t}} e^{-\frac{\alpha^2}{4t}}.
\end{equation}
Therefore, we can apply the Laplace transform in Eqn.~(\ref{screened}) such that
\begin{equation}
U\textsubscript{screened}(\mL,\vv) = \frac{e^{-\alpha \Vert\mL \vk + \vv\Vert}}{\Vert\mL \vk + \vv\Vert} = \frac{1}{\sqrt{\pi}} \int_0^\infty t^{-\frac{1}{2}} e^{-\Vert\mL\vk + \vv\Vert^2 t - \frac{\alpha^2}{4t}} dt.
\end{equation}
Then we obtain $A=1/\pi$, $B=\alpha^2$, $C=\frac{1}{2}$ and $D = 1/\sqrt{\pi}$ in Eqn.~(\ref{general}) to fit screened Coulomb potential into our potential summation method.

\section{Experimental Details}

\subsection{PotNet Implementation}
\label{model}

\begin{figure}[t]
    \centering
    \includegraphics[width=0.85\linewidth]{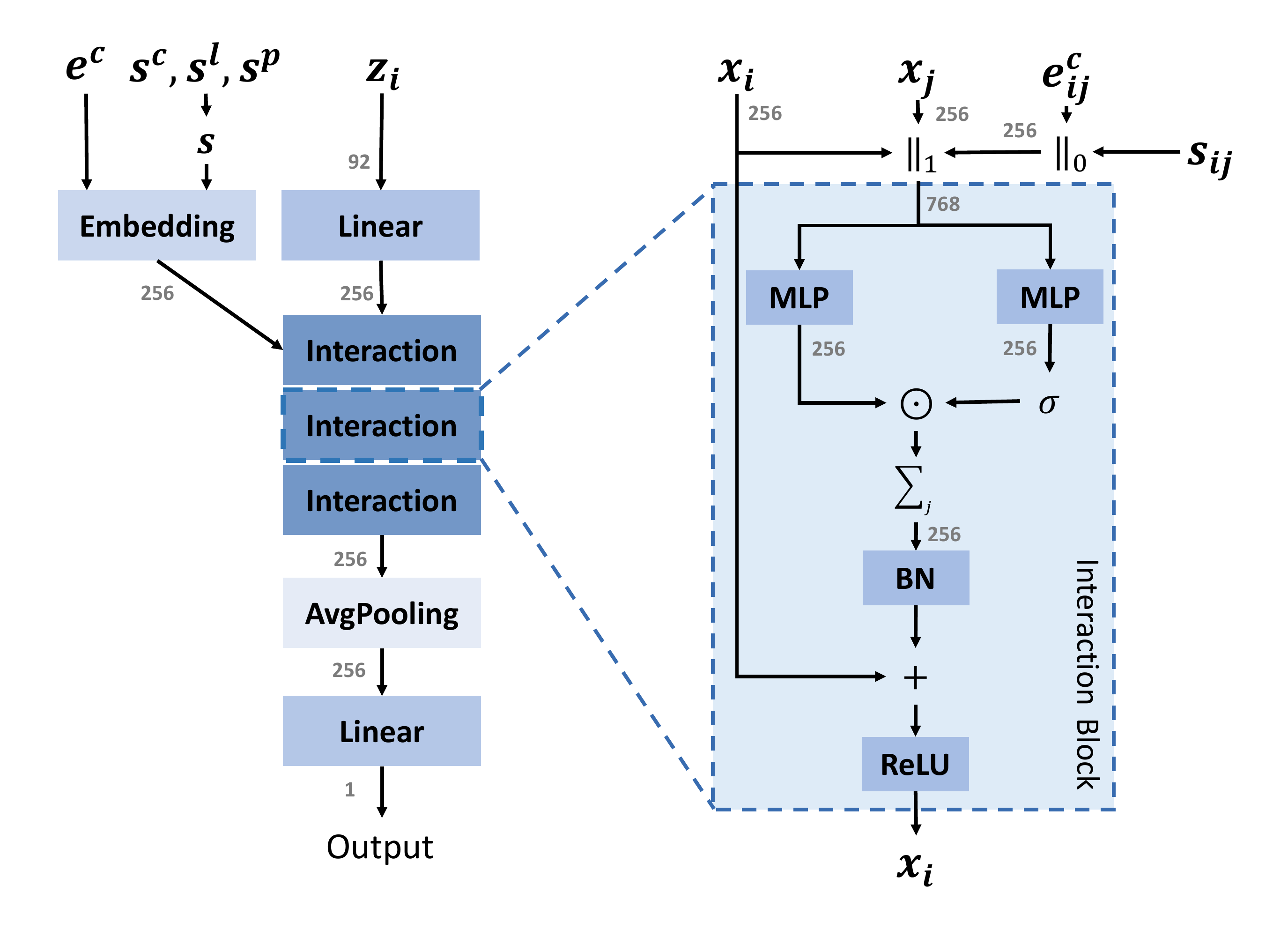}
    \caption{The developed network architecture for PotNet. The notations used in this figure are defined as follows: $\sigmoid$ denotes the sigmoid function; $\Vert_a$ represents the concatenation operation along the dimension $a$; $\odot$ signifies the Hadamard product, or element-wise multiplication of two matrices or tensors of the same dimensions; And $\sum_j$ represents the aggregation operation over the index $j$.
}
    \label{fig:model}
\end{figure}

The employed network architecture is shown in Fig.~\ref{fig:model}. Since our major contribution is to consider interatomic potentials and their complete form, we simply design our network architecture following the commonly used settings. Specifically, existing methods for 3D graphs~\citep{xie2018crystal,schutt2017schnet,klicpera2020directional,klicpera2020fast,gasteiger2021gemnet,schutt2021equivariant, wang2022comenet, liu2022spherical, yan2022periodic, wang2023learning} share a similar architecture, which usually contains an input block, an interaction block, and an output block. Without loss of generality, we take the updating process for node $i$ as an example to illustrate the network. 
\begin{itemize}

\item \textit{The Inputs} contain atomic features and potentials.
$\mathbf{z}_i$ is the 92-dimensional atomic feature for any atom $i$ following CGCNN~\citep{xie2018crystal}. Below the term $d$ denotes interatomic distances, which form an integral part of our potential features. As delineated in Sec.~\ref{sec:setup}, our computational model employs both local and global graphs. In the context of the local graph, only the Coulomb potential is considered, as per the details provided in the aforementioned Sec.~\ref{sec:setup}. On the other hand, the global graph encapsulates the summations of infinite potentials as its edge features. We consider all three categories of infinite potential summations in our model, namely Coulomb potentials, London dispersion potentials, and Pauli repulsion potentials. These are discussed in detail in Sec.~\ref{background:potentail}. The hyperparameter $\epsilon^{\prime}$ is employed to simplify the mathematical form of Coulomb potentials, expressed as $V\textsubscript{Coulomb}(\va,\vb)=-\frac{z_\va z_\vb e^2_0}{4\pi\epsilon_0 d(\va,\vb)}$, to $V\textsubscript{Coulomb}(\va,\vb)=-\epsilon^{\prime}/d(\va,\vb)$. This reduction is permissible since $e_0,\pi,\epsilon_0$ are all known constants, and $z_\va, z_\vb$ can be derived from atomic features. In the context of the local crystal graph, the Coulomb potentials are denoted as $\mathbf{e}^c = -\epsilon^{\prime\prime} /d$. For the infinite crystal graph, the summations of Coulomb potentials, London dispersion potentials, and Pauli potentials are denoted as $\mathbf{s}^c = -\sum_d \epsilon^{\prime}/d$, $\mathbf{s}^l = -\sum_d \epsilon/d^6$, and $\mathbf{s}^p= \sum_d e^{-\alpha d}$, respectively. We aggregate all three infinite features to derive the infinite features for the network, expressed as $s = s^c + s^l + k s^p$. In our experimentation, we fixed the values of the various parameters as follows: $\epsilon^{\prime\prime} = 0.75$, $\epsilon^{\prime} = 0.801$, $\epsilon = 0.074$, $\alpha = 3.0$, and $k=0.145$. The selection of these specific constants was the result of manual grid searching.

\item \textit{The Input Block} of our model consists of two primary components: a \textit{\textbf{Linear}} layer and an \textit{\textbf{Embedding}} layer. For every node $i$ in our model, we use the \textit{\textbf{Linear}} layer to generate a 256-dimensional vector. This vector serves as the input node features for the first interaction layer. Simultaneously, we apply an \textit{\textbf{Embedding}} layer for each edge in the model. This layer functions to map the Coulomb potentials and the summations of infinite potentials onto 256-dimensional embeddings. The Coulomb potentials undergo transformation using 256 Radial Basis Function (RBF) kernels, with the centers spanning a range from -4.0 to 4.0. Likewise, we transform the summations of infinite potentials using 64 RBF kernels, with centers also ranging from -4.0 to 4.0. Subsequently, we perform an up-projection on these transformed summations through a Multilayer Perceptron (MLP), resulting in a 256-dimensional output.

\item \textit{The Interaction Block} of our model comprises several \textit{\textbf{Interaction}} layers. Each of these layers dynamically updates the feature vector of a given node $i$, taking into account the features of its neighboring nodes as well as the potential embeddings of the connecting edges. More specifically, for any neighboring node $j$ of node $i$, the corresponding potential embeddings, denoted as $\mathbf{e}^c_{ij}$ and $\mathbf{s}_{ij}$, are generated by the \textit{\textbf{Embedding}} layer. These embeddings are initially concatenated along the edge dimension. Subsequently, the embeddings are concatenated with the node features $\mathbf{x}_i$ and $\mathbf{x}_j$ along the feature dimension. The principal interaction pattern of a layer parallels the pattern adopted in the CGCNN~\citep{xie2018crystal} as shown in the right side of Fig.~\ref{fig:model}. 

\item \textit{The Readout Block} of our model incorporates an \textit{\textbf{AvgPooling}} layer and a subsequent \textit{\textbf{Linear}} layer. Initially, the \textit{\textbf{AvgPooling}} layer is used to compile and aggregate features from all nodes within a graph. This aggregated feature set is then processed through the \textit{\textbf{Linear}} layer. The function of this layer is to map the hidden dimension, represented as a 256-dimensional vector, to a final scalar output.

\end{itemize}

\subsection{PotNet Improvement}
\label{sec:improve}
\begin{wraptable}[9]{r}{0.25\textwidth}
\vspace{-30 pt}
\begin{center}
\caption{PotNet improving techniques by periodic table information and transformer structure on JARVIS formation energy and training time. }\vspace{-0.1cm}
\label{tb:improvement}
\resizebox{0.25\textwidth}{!}{
\begin{tabular}{l|cc}
\toprule
Model & MAE & Time/Epoch \\ 
\midrule
& eV/atom & s \\
\midrule
PotNet & 0.0294 & 42 \\
PotNet-C  &	0.0293 & 42 \\
PotNet-T &	0.0290 & 50 \\
\bottomrule
\end{tabular}}
\end{center}
\end{wraptable} In order to enhance the efficacy of the model, we incorporated two strategies: the inclusion of periodic table information and the application of a transformer structure for infinite potential features. 

In our model, the charge information is inferred through atomic numbers, which could potentially introduce inaccuracies. To address this, we explicitly encode ten classes that are distinguished based on the categories of elements in the periodic table: alkali metals, alkaline earth metals, transition metals, post-transition metals, metalloids, reactive nonmetals, noble gases, lanthanides, and actinides. This encoding facilitates the model's learning of atomic properties. We have denoted this improved network model as PotNet-C.

Furthermore, we incorporated a transformer operation~\citep{graphormer, yan2022periodic} specifically applied to the infinite potential summation. This operation was particularly apt as the infinite potential summation is based on a fully-connected graph, an ideal fit for the transformer model. The transformer node output was then added to the local graph node output. This enhanced network structure, combined with the prior periodic table information, is referred to as PotNet-T.

As illustrated in Table.~\ref{tb:improvement}, PotNet-C displays performance comparable to that of the original PotNet, while PotNet-T exhibits superior results to both PotNet and PotNet-C. However, it should be noted that the computational efficiency of PotNet-T is lower, with only a modest increase in performance.

\subsection{Dataset Information}
\label{sec:dataset}

\begin{wraptable}[6]{r}{0.3\textwidth}
\vspace{-40 pt}
\begin{center}
\caption{Dataset information on JARVIS and the Materials Project.}\vspace{-0.2cm}
\label{tb:dataset}
\resizebox{0.3\textwidth}{!}{
\begin{tabular}{l|cc}
\toprule
Information & JARVIS & MP \\ 
\midrule
Size&	55722 &	69239\\
Mean Atom Numbers &	10.1 &	29.9\\
Mean Cell Length &	5.95$\AA$ &	8.02$\AA$\\
Minimum Cell Length &	0.99$\AA$&	1.78$\AA$\\
\bottomrule
\end{tabular}}
\end{center}
\end{wraptable} In Table~\ref{tb:dataset}, we detail the fundamental attributes of two distinct datasets: the Materials Project (MP) and the JARVIS 3D dataset. The considered characteristics comprise the size of each dataset, the average number of atoms per cell, the mean cell lengths, and the minimum cell length. As reflected in Table~\ref{tb:dataset}, the shortest cell length ranges approximately from 1$\AA$ to 2$\AA$. Many methods predicting crystal properties~\cite{xie2018crystal, schutt2017schnet, MegNet, ALIGNN} typically opt for a cutoff of either 4$\AA$ or 8$\AA$. This range, on average, only encompasses one-hop neighbors of a unit cell, and, at its most extensive, eight-hop neighbors.

\subsection{Full Ablation on JARVIS Dataset}
\label{sec:full_ablation}
\begin{table*}[t]
\caption{Ablation on our method with/without infinite potential summations in terms of test MAE on JARVIS dataset. The best results are shown in \textbf{bold}.}
\label{tb:full}
\begin{center}
\begin{tabular}{l|ccccc}
\toprule
& Formation Energy & Bandgap(OPT) & Total energy & Bandgap(MBJ) & Ehull\\ 

\cmidrule(r){2-6}
Method & eV/atom  &  eV & eV/atom & eV & eV   \\
\midrule
PotNet w/o Infinite & 0.0301 & 0.134 & 0.033 & 0.294 & 0.072\\
PotNet   & \textbf{0.0294} & \textbf{0.127} & \textbf{0.032} & \textbf{0.272} & \textbf{0.055}\\
\bottomrule
\end{tabular}
\end{center}
\vskip -0.25in
\end{table*}

In addition to the results shown in the main body, we conduct a thorough ablation study of the infinite potential summations, employing the JARVIS dataset for this analysis. As highlighted in Table.~\ref{tb:full}, the incorporation of infinite summation features contributes to consistent performance augmentation across all metrics evaluated in this study. Remarkably, a substantial enhancement is observed in the predictive accuracy associated with the BandGap (MBJ) and Ehull properties.

\subsection{Cutoff Experiments}
\label{sec:additional}

\begin{wraptable}[12]{r}{0.3\textwidth}
\vspace{-25 pt}
\begin{center}
\caption{Experiments with varied cutoffs on JARVIS formation energy and dataset preprocessing time.}\vspace{-0.2cm}
\label{tb:additional}
\resizebox{0.3\textwidth}{!}{
\begin{tabular}{l|cccc}
\toprule
Cutoff & SchNet & GATGNN & PotNet & Time \\ 
\midrule
\AA & eV/atom& eV/atom & eV/atom & s\\
\midrule
4&	0.052&	0.048 & 0.036 & 84\\
8&	0.045&	0.047 & 0.034 & 108\\
12&	0.045&	0.047 & 0.033 & 112\\
16&	0.045&	0.046 & 0.033  & 178\\
20&	0.044&	0.046 & 0.033 & 259\\
30&	0.042&	0.045 & 0.031 & 688\\
50&	Unstable&	0.043 & 0.030 & 3239 \\
\bottomrule
\end{tabular}}
\end{center}
\end{wraptable} In this section, we delve deeper into the investigation of two GNN approaches with differing cutoff values, with the aim of directly assessing the importance of complete interatomic potentials. The selection of these two methods is informed by their wide usage and the fact that they do not require additional graph construction techniques apart from radius crystal graph construction.

The first approach involves the application of the conventional GNN methodology, SchNet~\citep{schutt2017schnet}. The second approach, GATGNN~\citep{GATGNN}, incorporates the use of global attention. Additionally, we have incorporated our model without the infinite potential summation and local Coulomb potential into the study as the third approach. It should be noted that as the cutoff value increases, the feasibility of conducting training experiments diminishes due to the increase in time complexity. To facilitate training, we set the maximum number of an atom's neighboring atoms to 16.

We conduct training and testing on these three methodologies using the JARVIS 3D dataset, maintaining the same dataset settings as in Matformer~\citep{yan2022periodic}. We present the results pertaining to the formation energy in Table~\ref{tb:additional}. As indicated in Table~\ref{tb:additional}, it can be inferred that an increase in the cutoff value correlates with an enhancement in the performance of the three methodologies. However, at a cutoff value of 50$\AA$, the training of SchNet encounters a gradient explosion, rendering its final results unavailable. This issue may potentially stem from the modeling capacity of SchNet.

Further, we demonstrate the preprocessing time required for the entire JARVIS dataset, which comprises 55,722 crystals, in relation to different cutoff values in the fourth column of Table~\ref{tb:additional}. It becomes evident that as the cutoff value increases, the preprocessing time escalates to an unmanageable extent.

\section{Linear Energy Modeling using Infinite Potential Summation}
\label{born-lande}

\begin{wrapfigure}{r}{0.3\textwidth} 
\vspace{-10pt}
  \begin{center}
    \includegraphics[width=0.3\textwidth]{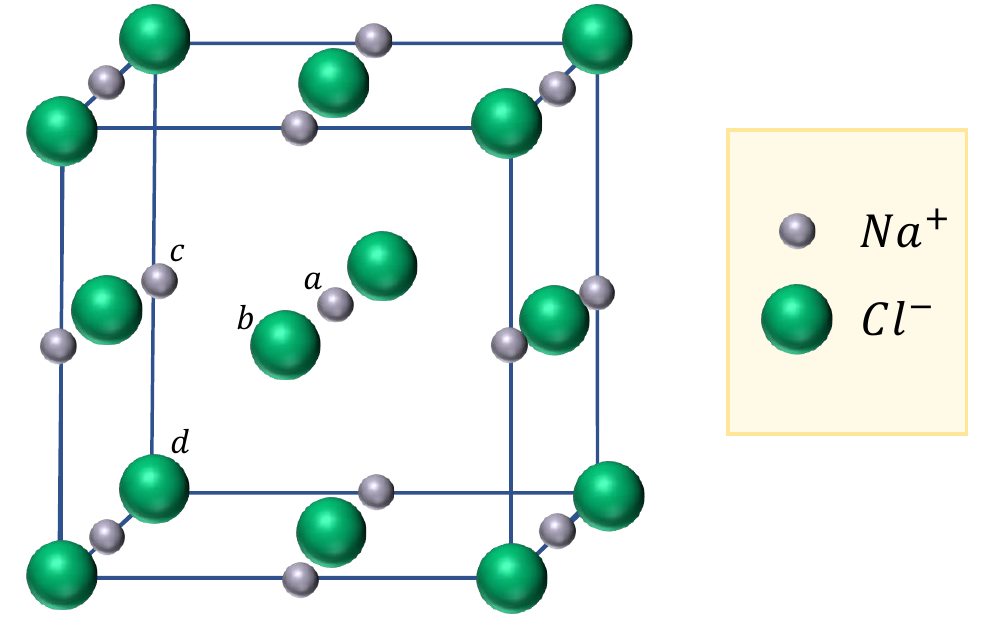}
    \vspace{-20pt}

    \caption{Crystal structure of $NaCl$.}
    \label{fig:nacl}
  \end{center}
  \vspace{-20pt}
\end{wrapfigure} In this section, we provide examples of calculations where the total energies of these materials can be directly approximated by linear combinations of infinite potential summations. These are special cases of Eqn.~(\ref{eqn:graph_eam}) with a linear embedded function $G$. Specifically, we evaluate the total energy per atom of $NaCl$ and two other materials ($MgO$, $LiF$) whose crystal structures are similar to $NaCl$. Since they are pure ionic crystals and Coulomb interactions dominate the system, we first consider their total electrostatic energy, i.e., the Coulomb potential summations. 

Inspired by analytic continuation as discussed in Appendix~\ref{sec:zeta}, we can approximate the total energy per atom of $NaCl$ by analytically continued infinite potential summations. The crystal structure of $NaCl$ is shown in Fig.~\ref{fig:nacl}. Due to the symmetry of the $NaCl$ cell, we only involve atoms $a,b,c,d$ in our calculation. Here, atom $a$ represents the body center $Na^+$, atom $b$ represents the face center $Cl^-$, atom $c$ represents the edge center $Na^+$, and atom $d$ represents the corner $Cl^-$. Given $N_A$ as Avogadro constant, $r_0$ as the minimum distance between $Na$ and $Cl$, $z_{Na}, z_{Cl}$ as charges of $Na^+$ and $Cl^-$, $e_0$ as the elementary charge constant, and $\epsilon_0$ as the permittivity constant of free space, the total energy of $Na^+$ is approximated by the total Coulomb interactions with atom $a$ such that
\begin{equation}
\label{cal}
\begin{aligned}
E_{Na} &= -N_A \bigg[1\cdot \frac{z_{Na}z_{Na}e^2_0}{4\pi \epsilon_0}\sum_{\vu\in\mA_\va,\vu\ne\va}\frac{1}{d(\va,\vu)}+ \frac{1}{2} 6 \cdot \frac{z_{Na}z_{Cl}e^2_0}{4\pi \epsilon_0}\sum_{\vu\in\mA_\vb}\frac{1}{d(\va,\vu)} \\
&\qquad + \frac{1}{4} 12 \cdot \frac{z_{Na}z_{Na}e^2_0}{4\pi \epsilon_0}\sum_{\vu\in\mA_\vc}\frac{1}{d(\va,\vu)} + \frac{1}{8} 8 \cdot \frac{z_{Na}z_{Cl}e^2_0}{4\pi \epsilon_0}\sum_{\vu\in\mA_\vd}\frac{1}{d(\va,\vu)} \bigg] \\
&= -N_A \bigg[\frac{z_{Na}z_{Na}e^2_0}{4\pi \epsilon_0 r_0}\sum_{\vu\in\mA_\va,\vu\ne\va}\frac{1}{\tilde{d}(\va,\vu)}+ 3 \cdot \frac{z_{Na}z_{Cl}e^2_0}{4\pi \epsilon_0 r_0}\sum_{\vu\in\mA_\vb}\frac{1}{\tilde{d}(\va,\vu)} \\
&\qquad + 3 \cdot \frac{z_{Na}z_{Na}e^2_0}{4\pi \epsilon_0 r_0}\sum_{\vu\in\mA_\vc}\frac{1}{\tilde{d}(\va,\vu)} +\frac{z_{Na}z_{Cl}e^2_0}{4\pi \epsilon_0 r_0}\sum_{\vu\in\mA_\vd}\frac{1}{\tilde{d}(\va,\vu)} \bigg] \\
&= -\frac{N_A |z_{Na}||z_{Cl}|e^2_0}{4\pi \epsilon_0 r_0} \bigg[\tilde{\mathcal{S}}(\va, \va) - 3\cdot\tilde{\mathcal{S}}(\va, \vb) + 3 \cdot\tilde{\mathcal{S}}(\va, \vc) - \tilde{\mathcal{S}}(\va, \vd)  \bigg]\\
&\approx -\frac{N_A |z_{Na}||z_{Cl}|e^2_0}{4\pi \epsilon_0 r_0} \bigg[-1.41864874 - 3 \cdot (-0.04796615) + 3 \cdot (-0.29126077) \\
&\qquad - (-0.40096799)  \bigg]\\
&\approx - \frac{N_A|z_{Na}||z_{Cl}|e^2_0}{4\pi \epsilon_0 r_0} \cdot (-1.7475646) \\
&\approx 8.81 eV,
\end{aligned}
\end{equation}
where $d(\va, \vu)$ is the distance between atom $\va$ and $\vu$, $\tilde{d} = d / r_0$ is the normalized distance, $\tilde{\mathcal{S}}$ is the infinite potential summation with $\tilde{d}$, approximated by our algorithm in Sec.~\ref{algorithm}, $\mA_\va$ denotes the set of atoms containing atom $\va$ and all its repetitions, and the coefficients $1, \frac{1}{2}6, \frac{1}{4}12, \frac{1}{8}8$ denote the fraction of atoms in a unit cell. We finally obtain a constant $-1.7475646$ and the total electrostatic energy approximation $8.81eV$ from our infinite potential summations. In fact, this constant $-1.7475646$ is exactly the famous Madelung constant $M$~\citep{borwein1985convergence} of $NaCl$. To obtain a more accurate total energy result by considering an additional repulsion term, we can derive the calculation result in Eqn.~(\ref{cal}) to the famous Born-Land\'e equation~\citep{BornConj}
\begin{equation}
\label{lande}
E = -\frac{N_A |z^+||z^-|e^2_0 M}{4\pi \epsilon_0 r_0} (1-\frac{1}{n}),
\end{equation}

where $z^+, z^-$ are the charges of cation and anion, $M$ is the Madelung constant computed from Coulomb potential summations, and $n$ is the Born exponent measuring the effect of repulsion. Choosing $n=9$, we can obtain an approximation for the total energy of $NaCl$ of 7.84 eV. Similarly, we also apply Eqn.~(\ref{lande}) to $MgO$ and $LiF$ to approximate the total energy per atom of these crystals. 

\begin{wraptable}[7]{r}{0.4\textwidth}
\vspace{-15 pt}
\begin{center}
\caption{Total energy per atom approximation of $NaCl$, $MgO$ and $LiF$.}\vspace{-0.3cm}
\label{tb:nacl}
\resizebox{0.4\textwidth}{!}{
\begin{tabular}{l|cc|cc}
\toprule
Formula & $r_0$ & $n$ & Ground Truth  & Eqn.~(\ref{lande})  \\ 
\midrule
$NaCl$ & 282 pm & 9 & 8.15 eV & 7.84 eV\\
$MgO$ & 210 pm & 6 & 39.33 eV & 39.45 eV \\
$LiF$ & 201 pm & 7 & 10.67 eV & 10.60 eV\\
\bottomrule
\end{tabular}}
\end{center}
\end{wraptable} We show these approximations in Table~\ref{tb:nacl} and it can be noticed that these approximations already give rough results compared to the ground truth total energy. This implies that our features can serve as a good starting point for machine learning models to learn the ground truth energy. Apparently, previous methods cannot achieve this due to the lack of such informative features. It is worth noting that the Madelung constant is typically unknown because those coefficients for the infinite potential summations depend on the charge distribution in the system, which we do not know at the beginning. Also, we already mention that these crystals are special cases of Eqn.~(\ref{eqn:graph_eam}) with a linear embedded function $G$, while $G$ is typically a nonlinear function~\citep{daw1984embedded}. Therefore, the network serves the purpose of learning those coefficients to learn the Madelung constant and providing nonlinearity.





\end{document}